\documentclass[12pt]{iopart}
\usepackage{graphicx}
\eqnobysec

\newcommand{\beq}{\begin{equation}}
\newcommand{\beqa}{\begin{eqnarray}}
\newcommand{\eeq}{\end{equation}}
\newcommand{\eeqa}{\end{eqnarray}}

\newcommand{\bigmean}[1]{\left\langle#1\right\rangle}
\renewcommand{\c}{{\rm c}}
\newcommand{\ceff}{{\rm c,eff}}
\renewcommand{\d}{{\rm d}}
\newcommand{\ds}{\displaystyle}
\newcommand{\eps}{\varepsilon}
\newcommand{\erfc}{\mathop{\rm erfc}\nolimits}
\newcommand{\euler}{{\gamma_{\scriptscriptstyle{\rm E}}}}
\newcommand{\frad}[2]{\ds{\frac{#1}{#2}}}
\newcommand{\haut}{\rule[-13pt]{0pt}{34pt}}
\newcommand{\lab}[1]{_{(#1)}}
\renewcommand{\max}{{\rm max}}
\newcommand{\mean}[1]{\langle#1\rangle}
\renewcommand{\min}{{\rm min}}
\newcommand{\prob}[1]{\mathop{\rm Prob}\nolimits\{#1\}}
\newcommand{\treshaut}{\rule[-23pt]{0pt}{54pt}}

\newcommand{\w}{\widehat}
\newcommand{\A}{{(\mathrm{A})}}
\newcommand{\B}{{(\mathrm{B})}}
\newcommand{\BA}{{\mathrm{BA}}}
\newcommand{\Iinid}{I_{\rm inid}}
\newcommand{\Int}{\mathop{\rm Int}\nolimits}
\renewcommand{\L}{{\cal L}}
\newcommand{\RD}{{\mathrm{RD}}}

\begin{document}

\title{On leaders and condensates in a growing network}

\author{C Godr\`eche and J M Luck}

\address{Institut de Physique Th\'eorique, IPhT, CEA Saclay,
and URA 2306, CNRS, 91191~Gif-sur-Yvette cedex, France}

\begin{abstract}
The Bianconi-Barab\'asi model of a growing network is revisited.
This model, defined by a preferential attachment rule
involving both the degrees of the nodes and their intrinsic fitnesses, has the
fundamental property to undergo a phase transition to a condensed phase
below some finite critical temperature, for an appropriate choice of the distribution of fitnesses.
At high temperature it exhibits a crossover to the Barab\'asi-Albert model,
and at low temperature, where the fitness landscape becomes very rugged,
a crossover to the recently introduced record-driven growth process.
We first present an analysis of the history of leaders,
the leader being defined as the node with largest degree at a given time.
In the generic finite-temperature regime, new leaders appear endlessly,
albeit on a doubly logarithmic time scale, i.e., extremely slowly.
We then give a novel picture for the dynamics in the condensed phase.
The latter is characterized by an infinite hierarchy of condensates,
whose sizes are non-self-averaging and keep fluctuating forever.
\end{abstract}


\eads{\mailto{claude.godreche@cea.fr},\mailto{jean-marc.luck@cea.fr}}

\maketitle

\section{Introduction}
\label{intro}

Growing network models with preferential attachment,
such as the Barab\'asi-Albert (BA) model~\cite{ba}, have become increasingly popular.
They bring a natural explanation for the scalefreeness observed in complex networks,
either natural or man made~\cite{abrmp,doro,blmch,cup,new}:
most networks exhibit a broad degree distribution falling off as a power law
($f_k\sim k^{-\gamma}$).
These model systems also provide a natural setting
to investigate other, more refined, features of networks
and other random growing structures.
In this work we focus our attention onto {\it leaders} and {\it lead changes}.
Luczak and Erd\"os~\cite{el} already describe as `a kind of a race'
the growth of connected components
in the Erd\"os-R\'enyi model for random graphs~\cite{er},
and refer to the largest component as the `leader'.
This terminology was then introduced in the physics literature
by Krapivsky and Redner~\cite{lkr},
in concomitance with the growing interest in social networks.
As suggested by these authors, since the degree of a node
may quantify the wealth of a corporation or the popularity of a person,
it is natural to investigate questions such as:\\
\noindent{\it How does the identity of the leader (the node with largest degree) change in the course of time?}
{\it What is the probability that a leader retains the lead as a function of time?}

The present work is devoted to an investigation of questions of this kind
in the Bianconi-Barab\'asi (BB) fitness model~\cite{bb1,bb2}.
This model has a natural temperature parameter~$T$.
It interpolates between the BA model~\cite{ba} ($T=\infty$)
and the record-driven (RD) growth model~\cite{R} ($T=0$).
It exhibits a condensation transition at some finite critical temperature $T_\c$,
for an appropriate choice of the distribution of fitnesses.
Our main goal is to extend to the full BB model
the analysis of the statistics of leaders and lead changes,
respectively performed in~\cite{R} and~\cite{lkr,L} for the RD and BA models.
We shall consider the case where the model has a condensation transition,
in order to investigate the differences in behavior between the fluid phase ($T>T_\c$)
and the condensed phase ($T<T_\c$).
Besides the interest in leaders {\it per se},
the present study will also shed new light onto the dynamics
of the condensed phase of the BB model, a subject so far only tackled in~\cite{fb}.

The BB model is a growing network model with preferential attachment.
A new node enters the network at each time step.
The node entering at time~$n$ is labeled by its index $n$.
It attaches to a single earlier node ($i=1,\dots,n-1$) with probability
\beq
p_{n,i}=\frac{\eta_i\,k_i(n-1)}{Z(n-1)}.
\label{pnidef}
\eeq
The attachment probabilities $p_{n,i}$ combine both a {\it fit-get-richer} feature,
through the intrinsic fitnesses $\eta_i$ of the nodes,
and a {\it rich-get-richer} feature, through their dynamical degrees $k_i(n-1)$.
More precisely:

\noindent {\bf (i)}
The fitness $\eta_i$
is assigned to node $i$ once for all and modeled as a quenched random variable.
Fitnesses are conveniently parametrized as activated variables:
\beq
\eta_i=\exp(-\eps_i/T),
\label{activ}
\eeq
temperature $T$ providing a measure of the ruggedness of the fitness landscape,
whereas the activation energies $\eps_i$ are independent and identically distributed
(i.i.d.) random variables, drawn from some temperature-independent distribution.
Nodes $i$ with a low energy $\eps_i$ have a high fitness $\eta_i$,
and therefore a larger chance to attract new connections.
The lower bound of energies is set for convenience to $\eps=0$.
The behavior of the density $\rho(\eps)$ of the distribution for $\eps\to0$
turns out to play a central r\^ole in the model.
We assume that $\rho(\eps)$ starts as a power law:
\beq
\rho(\eps)\sim\eps^{\theta-1}\qquad(\eps\to0),
\label{rhopower}
\eeq
with exponent $\theta>0$.
In numerical simulations we shall use the triangular distribution
\beq
\rho(\eps)=2\eps\qquad(0<\eps<1),
\label{rhotri}
\eeq
which has $\theta=2$.

\noindent {\bf (ii)}
The degree $k_i(n-1)$ is
the number of connections of node $i$ just before node $n$ enters the network.
Nodes with a high degree, i.e., which are already well connected,
have a larger chance to attract new connections.

\noindent {\bf (iii)}
Finally, the partition function $Z(n-1)$ in the denominator of~(\ref{pnidef})
ensures that the attachment probabilities are normalized:
\beq
Z(n)=\sum_{i=1}^n\eta_i\,k_i(n).
\label{zdef}
\eeq

As done in our previous studies~\cite{L,D}, we will hereafter consider two different initial conditions in parallel:

\noindent {\it Case~A.}
Node~1 appears at time $n=1$ with degree $k_1(1)=0$.
At time $n=2$ node~2 attaches to node~1, so that $k_1(2)=k_2(2)=1$.
At time $n=3$ node~3 can attach either to node~1 or to node~2.
Hereafter we make the choice of attaching node~3 to node~1,
obtaining thus $k_1(3)=2$, whereas $k_2(3)=k_3(3)=1$.

\noindent {\it Case~B.}
Node~1 appears at time $n=1$ with degree $k_1(1)=1$.
This amounts to saying that the first node is connected to a root,
which does not belong to the network.
It is natural to represent this connection by half a link.
At time $n=2$ node~2 attaches to node~1.
We thus have $k_1(2)=2$ and $k_2(2)=1$.

In both cases the network has the topology of a tree.
The sum of the node degrees at time $n$
equals twice the number of links $L(n)$ in the network:
\beq
\sum_{i=1}^n k_i(n)=2L(n),
\label{bonds}
\eeq
with $2L^\A(n)=2n-2$ and $2L^\B(n)=2n-1$.
Here and in the following, the superscripts~$\A$ and $\B$ denote a result
which holds for a prescribed initial condition, i.e., Case~A or Case~B.
We have therefore $\mean{k}^\A(n)=2-2/n$ and $\mean{k}^\B(n)=2-1/n$,
which both yield
\beq
\mean{k}=2
\label{ksum}
\eeq
in the thermodynamic limit, as expected for a tree topology.

Owing to its temperature parameter $T$, the BB model realizes
a continuous interpolation between two limiting models.

\begin{itemize}
\item
The Barab\'asi-Albert (BA) model~\cite{ba}, for $T=\infty$.

In this limit we have $\eta_i=1$ for all nodes~$i$,
so that node fitnesses do not play any r\^ole.
The BA model is thus recovered.
The attachment probability $p_{n,i}$ is proportional to the degree $k_i(n-1)$.
The partition function reads $Z(n)=2L(n)$ (see~(\ref{bonds})).

\item
The record-driven (RD) growth model~\cite{R}, for $T=0$.

In this limit, the growth of the network is record-driven.
The current record node, i.e., the node whose fitness is the largest,
attracts all the new connections, until it is outdone by the next record.

\end{itemize}
Let us finally introduce some definitions used in the present work.

\medskip
\noindent (a) {\it Leaders and lead changes}

A {\it co-leader} at time $n$ is any node whose degree is equal to the largest degree
\beq
k_\max(n)={\rm max}(k_1(n),\dots,k_n(n)).
\label{kmaxdef}
\eeq
The {\it leader} at time $n$ is the node
among the co-leaders whose degree reached the value $k_\max(n)$ {\it first}.
We denote by~$I(n)$ the {\it index of the leader} at time $n$.
Initial conditions are such that the first node is the first leader.
There is a {\it lead change} at time $n$
if the leader at time $n$ is different from that at time~$n-1$.
We call a {\it lead} any period of time between two lead changes.
We denote by $\L(n)$ the {\it number of leads} up to time $n$.
In other words, the number of lead changes up to time $n$ is $\L(n)-1$.
Some lead changes bring to the lead a node that has already been the leader in the past,
whereas some other changes promote a newcomer.
We denote by $D(n)$ the {\it number of distinct leaders} up to time $n$.
We also define the {\it lead persistence probability} $S(n)$
as the probability that there is a single leader up to time $n$.

\medskip
\noindent (b) {\it Records}

The {\it record node} at time $n$ is the node whose energy (fitness)
is the smallest (largest) met up to time $n$.
A node is said to be a record if it belongs to the series of records,
i.e., if it has been the record during some time lapse in the past.
We denote by $\Pi(n)$ {\it the probability that the leader at time $n$ is a record}.

The setup of this paper is as follows.
In Section~\ref{continuum} we revisit the continuum mean-field-like approach
of~\cite{bb1,bb2},
emphasizing in particular finite-size effects near the critical point.
Section~\ref{leaders} is devoted to the history of leaders and lead changes
during a typical instance of the growth of the network,
and Section~\ref{snapshot} to an investigation of the degree statistics in the condensed phase.
A brief discussion followed by a summary are given in Section~5.
An example of extreme-value statistics for i.n.i.d.~random variables
and a few complements on the record-driven growth process
are respectively exposed in Appendix~A and Appendix~B.

\section{The continuum formalism revisited}
\label{continuum}

The main features of the model,
and in particular the possible occurrence of a condensation transition,
have been studied in the original works of Bianconi and Barab\'asi~\cite{bb1,bb2}.
They use a continuum mean-field-like formalism,
where both the discrete structure of the model and all fluctuations are neglected.
In this section, we revisit this analysis
in order to unveil some features of the model untouched so far,
such as the importance of finite-size effects near the critical point,
or the possible occurrence of an infinite hierarchy of condensates in the low temperature phase.

\subsection{Phase diagram}
\label{diagram}

Our starting point is the following exact recursion equation for the mean degree:
\beq
\mean{k_i(n)}-\mean{k_i(n-1)}=\mean{p_{n,i}}\qquad(n>i),
\label{eqmean}
\eeq
where brackets denote an average over the history of the network up to time~$n$,
i.e., both over the stochastic growth process
and over the quenched random fitnesses.
In the spirit of the continuum formalism,
i.e., neglecting all fluctuations and treating time $n$ as continuous,
we approximate the above equation as (see~(\ref{pnidef})):
\beq
\frac{\partial k_i}{\partial n}\approx\frac{\eta_ik_i}{Z(n)}.
\label{park}
\eeq
We furthermore estimate the partition function in the denominator
as $Z(n)\approx Cn$, with
\beq
C=\mean{\eta k}
\label{csum}
\eeq
(see~(\ref{zdef})).
The differential equation~(\ref{park}) can then be integrated as
\beq
k_i(n)\approx\left(\frac{n}{i}\right)^{\eta_i/C}.
\label{kin}
\eeq
This result is the cornerstone of the continuum approach.

The mean degree $\bar{k}(\eta)$ of a node with fitness $\eta$
can then be estimated by uniformly averaging~(\ref{kin}) over the index $i$:
\beq
\bar{k}(\eta)\approx\frac{1}{n}\sum_{i=1}^n\left(\frac{n}{i}\right)^{\eta/C}
\approx\frac{C}{C-\eta},
\label{keta}
\eeq
where the rightmost expression is obtained by replacing the sum by an integral.
The mean degree $\mean{k}$ can then be estimated
by averaging~(\ref{keta}) over the distribution of the fitness $\eta$.
We thus obtain
\beq
\mean{k}=1+K(T,C),
\label{kmean}
\eeq
and similarly
\beq
\mean{\eta k}=C\,K(T,C),
\label{cmean}
\eeq
with~\cite{bb1,bb2}
\beq
K(T,C)=\bigmean{\frac{\eta}{C-\eta}}
=\int_0^\infty\frac{\rho(\eps)\,\d\eps}{C\,\e^{\eps/T}-1}\qquad(C>1)
\eeq
(see~(\ref{activ})).
The equalities~(\ref{ksum}) and~(\ref{csum}) consistently yield the condition
\beq
K(T,C)=1,
\label{bbc}
\eeq
which determines the parameter $C$ as a function of temperature.

As said earlier,
the main feature of the BB model is the possible occurrence of a condensation transition
at some finite temperature $T_\c$~\cite{bb1,bb2},
according to the value of the exponent~$\theta$ of the distribution
of the energies $\eps_i$ (see~(\ref{rhopower})).

\noindent $\bullet$
If $\theta\le1$, we have $K(T,C)\to\infty$ as $C\to1^+$ for any finite temperature.
The condition~(\ref{bbc}) defines a regular function $C(T)$,
decreasing from $C(\infty)=2$ to $C(0)=1$.

\noindent $\bullet$
If $\theta>1$, the integral $K(T,C)$ is convergent as $C\to1^+$.
The limit $K(T,1)$ is a decreasing function of temperature,
such that $K(T,1)\to\infty$ as $T\to\infty$ and $K(T,1)\to0$ as $T\to0$.
There is a unique {\it critical temperature} $T_\c$ such that
\beq
K(T_\c,1)=1.
\label{bbtc}
\eeq
The condition $\theta>1$ is somewhat similar to the condition $d>2$ for the occurrence
of a Bose-Einstein condensation in a free Bose gas.

\noindent $\star$ In the high-temperature (fluid) phase ($T>T_\c$),
the function $C(T)$ is regular and decreasing from $C(\infty)=2$ to $C(T_\c)=1$
at the critical point.

\noindent $\star$ In the low-temperature (condensed) phase ($T<T_\c$),
$C(T)=1$ stays equal to its critical value,
whereas the r.h.s.~of~(\ref{kmean}) keeps varying with temperature.
This implies that a macroscopic fraction of the degrees
escapes the sum rules~(\ref{ksum}) and~(\ref{csum}).
The analogy with the Bose-Einstein mechanism
suggests that a condensation takes place~\cite{bb2}.
Assume that some finite number $J$ of nodes, labeled $i_j$ for $j=1,\dots,J$,
are condensates,
in the sense of having macroscopic degrees $k_{i_j}\approx R\lab{j}n$,
proportional to the total number of nodes $n$,
and fitnesses $\eta_{i_j}\approx1$ close to optimal.
Let $F=\sum_j R\lab{j}$ be the total size (i.e., degree fraction) of these condensates.
Equations~(\ref{kmean}) and~(\ref{cmean}) read
\beq
\mean{k}=1+K(T,C)+F=2,\qquad\mean{\eta k}=C\,K(T,C)+F=C,
\eeq
hence $C=1$ and
\beq
F=1-K(T,1).
\label{stotal}
\eeq

The total size $F$ of the condensates
is thus predicted to be a well-defined function of temperature,
increasing smoothly as temperature is lowered, from $F=0$ at the critical point
to $F=1$ in the $T\to0$ limit.
Let us emphasize that the continuum approach neither predicts
the number $J$ of condensates nor the individual sizes $R\lab{j}$ of the condensates.
The possibility of there being more than one condensate in the low-temperature phase
of the BB model is hinted at in~\cite{new}.
We shall investigate these matters in Section~\ref{snapshot}.

\subsection{Finite-size effects}
\label{fss}

We now turn to an investigation of finite-size effects near the critical temperature.
Let us have a closer look at the sum appearing in~(\ref{keta}).
It is of the form
\beq
S_n(\delta)=\sum_{i=1}^n i^{\delta-1},
\eeq
where the exponent $\delta=1-\eta/C$ is in the range $0\le\delta\le1$.
For fixed $\delta>0$, we can estimate $S_n(\delta)$ at large $n$
by replacing the sum by an integral.
We thus obtain $S_n(\delta)=n^\delta/\delta+\cdots$,
where the dots stand for a $\delta$-dependent correction of order unity.
For $\delta=0$, $S_n(0)$ is equal to $H_n$, the harmonic number of order $n$,
and therefore diverges logarithmically as $S_n(0)\approx\ln n+\euler$,
where $\euler$ is Euler's constant.
In the scaling region where $\delta$ is small and $n$ is large,
the sum behaves as
\beq
S_n(\delta)\approx\frac{\e^{\delta\ln n}-1}{\delta}.
\label{fsss}
\eeq
This expression interpolates between the two limiting forms seen above.
It also puts forward a time-dependent scale of $\delta$,
\beq
\delta_\star\approx\frac{1}{\ln n},
\label{delstar}
\eeq
to which we shall come back in Section~\ref{leaders}.
Inserting the scaling form~(\ref{fsss}) into~(\ref{keta}),
we obtain a finite-size correction to~(\ref{kmean}) of the form
\beq
\mean{k}\approx1+K(T,C)-\Delta K(T,C),\qquad\Delta K(T,C)
=\bigmean{\frac{n^{-(1-\eta/C)}}{1-\eta/C}}.
\label{kfss}
\eeq

In the high-temperature phase ($T>T_\c$), we have $C(T)>1$,
so that $\Delta K(T,C)\sim n^{-\omega(T)}$ falls off as a power law,
with a positive correction exponent
\beq
\omega(T)=1-1/C(T).
\eeq
Finite-size corrections for generic quantities obey the same power-law fall-off.
The well-known $n^{-1/2}$ scaling of finite-size corrections in the BA model
is recovered in the $T=\infty$ limit, since $C(\infty)=2$.

At the critical point, we have $C(T_\c)=1$,
so that the correction exponent $\omega(T_\c)$ vanishes,
and a more accurate treatment is needed.
Assuming $C=1$ exactly, and hence $T\approx T_\c$, we have $1-\eta/C\approx\eps/T_\c$.
As a consequence,
the expression~(\ref{kfss}) for $\Delta K(T_\c,1)$ is dominated by small values
of the activation energy $\eps$, and simplifies to
\beq
\Delta K(T_\c,1)\approx A\,T_\c\int_0^\infty\eps^{\theta-2}\,\e^{-(\eps\ln n)/T_\c}\,\d\eps
=\frac{A\,\Gamma(\theta-1)\,T_\c^\theta}{(\ln n)^{\theta-1}}.
\eeq

In order to investigate finite-size scaling around the critical point,
we define an effective time-dependent critical temperature $T_\ceff(n)=T_\c+\Delta T(n)$
by requiring that $C=1$, i.e., more precisely,
that the mean partition function equals
$\mean{Z(n)}=n$ at $T=T_\ceff(n)$.
The shift $\Delta T(n)$ provides a measure of the size of the critical region
for a large but finite time $n$.
It can be estimated by setting $\mean{k}=2$ in~(\ref{kfss}).
We thus obtain
\beq
\Delta T(n)\approx\frac{\Delta K(T_\c,1)}{\mu(T_\c)}
\sim\frac{1}{(\ln n)^{\theta-1}}.
\label{dtc}
\eeq
The denominator
\beq
\mu(T)=\frac{\partial K(T,1)}{\partial T}=\frac{1}{T^2}
\int_0^\infty\frac{\eps\,\rho(\eps)\,\d\eps}{(\e^{\eps/T}-1)^2}
\label{denopar}
\eeq
is indeed convergent whenever $\theta>1$.

We now give a few explicit formulas
in the case of the triangular distribution~(\ref{rhotri}).
This example of a distribution with $\theta=2$
will be used in numerical simulations in the following.
We have
\beq
K(T,C)=\int_0^1\frac{2\eps\,\d\eps}{C\,\e^{\eps/T}-1}
=2T\sum_{m\ge1}\frac{T-(m+T)\e^{-m/T}}{m^2C^m},
\eeq
so that
\beq
K(T,1)=\frac{\pi^2T^2}{3}-2T\sum_{m\ge1}\frac{(m+T)\e^{-m/T}}{m^2}
\eeq
and
\beq
\mu(T)=\frac{2\pi^2T}{3}-\frac{2}{T}\sum_{m\ge1}\frac{(m^2+2mT+2T^2)\e^{-m/T}}{m^2}.
\eeq
All the above series are exponentially convergent.
The condition~(\ref{bbtc}) yields
\beq
T_\c\approx0.711\,716\,308,\qquad\mu(T_\c)\approx1.896\,474\,189.
\label{tc}
\eeq

\begin{figure}
\begin{center}
\includegraphics[angle=-90,width=.5\linewidth]{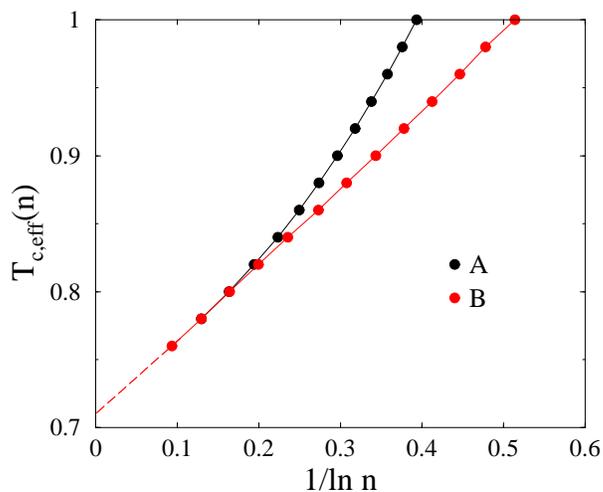}
\caption{\label{teff}
Effective temperature $T_\ceff(n)$ against $1/\ln n$.
Symbols: numerical data for both initial conditions.
Dashed line: extrapolation of the data for initial condition~B.}
\end{center}
\end{figure}

Figure~\ref{teff} shows numerical results illustrating
the slow logarithmic fall-off of the size of the critical region predicted by~(\ref{dtc}).
Here and throughout the following,
data are produced by means of a direct numerical simulation of the BB model
for the triangular distribution~(\ref{rhotri}).
Data are averaged over $10^5$ independent realizations
for each initial condition (A or~B).
The figure shows the effective temperature $T_\ceff(n)$ against $1/\ln n$.
The observed linear convergence corroborates the prediction~(\ref{dtc}).
The fit of the data for Case~B (dashed line) extrapolates to $T_\c\approx0.711$,
thus reproducing the critical point~(\ref{tc}) to three digits.

\section{The history of leaders and lead changes}
\label{leaders}

\subsection{The index of the leader: a variational approach}
\label{variational}

In order to estimate the index $I(n)$ of the leader at time $n$,
we start from the prediction~(\ref{kin}) of the continuum formalism
for the degree of node $i$ at time $n$, that we rewrite as
\beq
C\ln k_i(n)\approx Z_i=\e^{-\eps_i/T}\,\ln\frac{n}{i}\qquad(i=1,\dots,n).
\label{yin}
\eeq
Within this framework,
which neglects the noise involved in the growth process,
estimating the degree $k_\max(n)$ and the index $I(n)$ of the leader
amounts to studying the largest value $Z_\max$ of the $n$ random numbers $Z_i$
defined above and the index, hereafter denoted by $\Iinid$, where this largest value is reached.

In other words, we are facing a problem of extreme-value statistics
for independent but not identically distributed (i.n.i.d.) random variables.
The classical extreme-value theory for independent and identically distributed
(i.i.d.)~variables is well documented~\cite{evs}.
In contrast, little is known for i.n.i.d.~random variables
(see~\cite[Section~4.2]{cb} for a recent review).
The present problem provides an example of strongly i.n.i.d.~variables.
The support of the random variables, i.e., $0<Z_i<\ln(n/i)$,
indeed shrinks rapidly as a function of the index $i$.
The index~$\Iinid$ is therefore expected to be localized
in a relatively small range $w\ll n$ of values of $i$.

In order to estimate the localization range $w$,
we propose a variational approach,
designed for the study of strongly i.n.i.d.~random variables,
i.e., whose supports have a strong dependence on their index.
The predictions of the variational approach
will be tested against exact results for a specific example in Appendix~A.
This approach proceeds as follows.
Neglecting all numerical prefactors, we evaluate the index $\Iinid$ as~$w$.
The minimal energy $\eps_\min$ reached in this range
is evaluated as the smallest in a set of $w$ i.i.d.~variables~$\eps_i$.
According to a well-known argument of extreme-value statistics,~$\eps_\min$ is
such that the probability of having $\eps<\eps_\min$, i.e.,
$\eps_\min^\theta$ (see~(\ref{rhopower})), is of order $1/w$.
Putting both above estimates together, we obtain
\beq
Z_\max\sim\exp\left(-\frac{1}{T w^{1/\theta}}\right)\ln\frac{n}{w}.
\eeq
This expression has a non-trivial maximum as a function of $w$.
If $w$ is too large, the typical weight $\ln(n/w)$ is too small,
whereas if $w$ is too small, $\eps_\min$ is too large because of an insufficient sampling.
It is natural to estimate the localization range $w$ as the location of this maximum.
We thus obtain the following implicit equation for $w$:
\beq
w\sim\left(\frac{1}{T\theta}\ln\frac{n}{w}\right)^\theta.
\label{wimplicit}
\eeq

Consistently neglecting prefactors and subleading corrections,
we are left with the following prediction,
both for the index $\Iinid$ where the largest value $Z_\max$ is reached,
and for the index $I(n)$ of the leader in the original problem:
\beq
I(n)\sim\Iinid\sim w\sim\left(\frac{\ln n}{T}\right)^\theta.
\label{wsca}
\eeq
In turn $Z_\max$ and $k_\max$ can be estimated as
\beq
Z_\max\sim\ln\frac{n}{w},\qquad
k_\max(n)\sim\left(\frac{n}{w}\right)^{1/C}
\sim\left(n\left(\frac{T}{\ln n}\right)^\theta\right)^{1/C}.
\label{ysca}
\eeq
These scaling predictions are the main results of this section.
Let us note that the corresponding minimal energy
scales as $\eps_\min\sim T/\ln n$.
The associated value of the gap $\delta=1-\eta_\max\approx\eps_\min/T$,
i.e., $\delta\sim1/\ln n$, is consistent with the scale~(\ref{delstar}).

\begin{figure}
\begin{center}
\includegraphics[angle=-90,width=.5\linewidth]{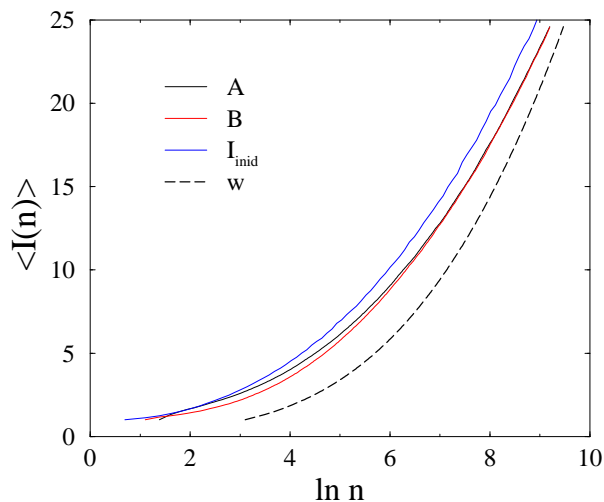}
\caption{\label{wtest}
Black and red: mean index $\mean{I(n)}$ of the leader
against $\ln n$ for $T=0.4$ and both initial conditions.
Blue: mean index $\mean{\Iinid}$ corresponding to the variables~$Z_i$.
Black dashed curve: localization range $w$ predicted by~(\ref{wimplicit}),
translated horizontally for readability.}
\end{center}
\end{figure}

Before analyzing the consequences of the above results,
in Figure~\ref{wtest} we present a check of the variational approach
against numerical data for a temperature $T=0.4$ in the condensed phase.
The black and red curves show data for the mean index $\mean{I(n)}$ of the leader
against $\ln n$ for initial conditions A and B.
Both essentially yield the same results, except at short times.
The blue curve shows data for the mean index $\mean{\Iinid}$
corresponding to the largest of the i.n.i.d.~variables $Z_i$ introduced in~(\ref{yin}).
The data have been obtained by averaging over $10^5$ independent samples
of these variables for each value of $n$.
The black dashed line shows the localization range $w$
as predicted by the implicit equation~(\ref{wimplicit}).
For clarity, $n$ has been multiplied by an arbitrary factor
(i.e., $\ln n$ has been translated).
The good agreement between $\mean{I(n)}$ and $\mean{\Iinid}$
demonstrates that the study of the variables $Z_i$ dictated by the continuum approach
captures the essential features of the growth of the degrees.
The qualitative agreement of the localization range~$w$
with all the data corroborates the validity of the variational approach.

The main prediction of the variational approach
is that the typical index of the leader at time $n$
grows as $I(n)\sim(\ln n)^\theta$ (see~(\ref{wsca})).
This slow growth is expected to hold irrespectively of temperature $T$,
i.e., both in the fluid phase $(T>T_\c)$ and in the condensed phase ($T<T_\c)$.

At variance with the index of the leader,
the growth law~(\ref{ysca}) of the degree $k_\max(n)$ of the leader
depends on the phase of the model through the dependence of $C$ itself on temperature.

\begin{itemize}

\item
In the fluid phase $(T>T_\c)$,
we obtain the subextensive power-law growth $k_\max\sim n^{1/C(T)}$,
up to a logarithmic correction.
The exponent $1/C(T)=1-\omega(T)$ coincides with the maximal growth exponent
of the expression~(\ref{kin}).
It is a decreasing function of temperature,
varying continuously between the BA exponent $1/2$ in the $T\to\infty$ limit
and the limit value 1 as $T\to T_\c^+$.

\item
Right at the critical temperature $(T=T_\c)$, the prediction~(\ref{ysca}) reads
\beq
k_\max\sim\frac{n}{(\ln n)^\theta}.
\label{kcsca}
\eeq
The degree of the leader is therefore only logarithmically subextensive.
The validity of this prediction is demonstrated in Figure~\ref{kc},
showing the ratio $n/\mean{k_\max}$ against $\ln n$.
The data for both initial conditions are well represented by a quadratic polynomial
(blue line), obtained by fitting the data for initial condition B,
and slightly translated for readability.

\begin{figure}
\begin{center}
\includegraphics[angle=-90,width=.5\linewidth]{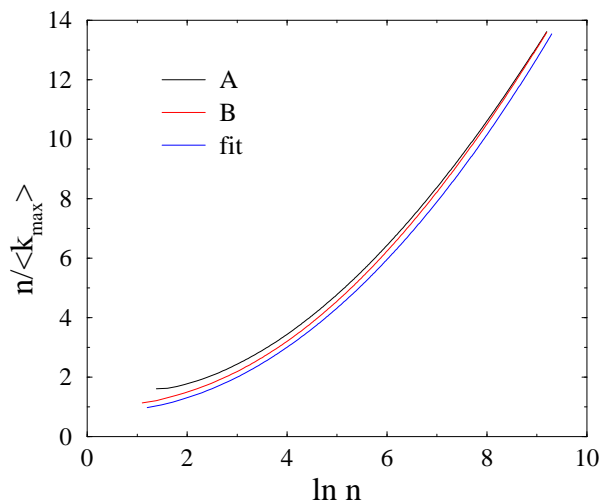}
\caption{\label{kc}
Black and red: ratio $n/\mean{k_\max}$ against $\ln n$
at the critical point $(T=T_\c)$, for both initial conditions.
Blue: fit by a quadratic polynomial, slightly translated for readability.}
\end{center}
\end{figure}

\item
In the condensed phase ($T<T_\c$), the expression~(\ref{ysca}) is of little use,
because of the presence of large finite-size corrections to $C(T)$.
The degree statistics in the condensed phase
will be investigated in Section~\ref{snapshot}.

\end{itemize}

\subsection{A typical history of leaders and lead changes}
\label{scenario}

The scenario we are proposing for the typical history of leaders and lead changes
is based on the following two ingredients.

First and most importantly, the index of the leader at time $n$
grows as $I(n)\sim(\ln n)^\theta$ (see~(\ref{wsca})),
irrespectively of temperature.
There is therefore an asymptotic decoupling of time scales between $n$ and $I(n)$.

\begin{figure}
\begin{center}
\includegraphics[angle=-90,width=.5\linewidth]{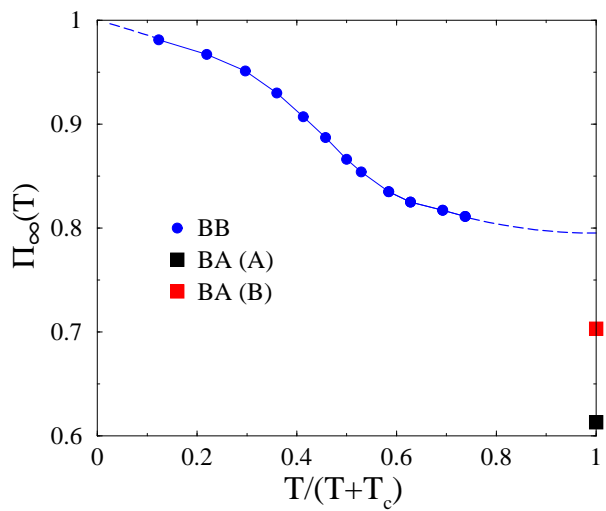}
\caption{\label{q}
Asymptotic probability $\Pi_\infty(T)$
for the leader to belong to the series of records, against $T/(T+T_\c)$.
Dashed lines: extrapolation of fits to the data.
Symbols on right axis: corresponding quantities in the BA model
with both initial conditions.}
\end{center}
\end{figure}

The second ingredient
is the high probability that the leader is a record.
The probability $\Pi(n,T)$ that the leader at time $n$ is a record,
in the sense that it belongs to the series of record nodes,
is observed to converge
to an asymptotic value $\Pi_\infty(T)$, irrespectively of the initial condition.
This limit value is plotted in Figure~\ref{q} against $T/(T+T_\c)$.
The crossovers to the BA and RD models at high and low temperatures
(see the next two sections)
prevent one from having accurate data at either end of the temperature range.
The asymptotic probability is observed to decrease as a function of temperature,
from $\Pi_\infty\to1$ in the $T\to0$ limit, through $\Pi_\infty(T_\c)\approx0.866$,
to a limit $\Pi_\infty\approx0.80$ as $T\to\infty$.
The symbols to the right show the corresponding quantities for the BA model,
i.e., $\Pi_\infty^\A\approx0.613$ and $\Pi_\infty^\B\approx0.703$.
The discrepancy between these numbers and the previous limit
is a manifestation of the discontinuous behavior to be described
in Section~\ref{highcross}.

Summarizing what we have learned so far about the leader in a large network
consisting of~$n$ nodes,
the index of the leader typically grows as $(\ln n)^\theta$,
whereas it has a high chance of being a record.
Therefore, as a rule of thumb:
{\it The Leader of Today is the Record of Old.}
The scenario for the history of leaders and lead changes to which we come up
is essentially insensitive to the value of temperature,
and is therefore expected to hold both in the fluid phase $(T>T_\c)$
and in the condensed phase ($T<T_\c)$.

Several consequences can be drawn from this scenario.
The typical number $D(n)$ of distinct leaders up to time $n$
is expected to grow (at most)
as the typical number of records up to time $I(n)\sim(\ln n)^\theta$, i.e.,
$D(n)\sim\ln I(n)$, or
\beq
D(n)\sim\theta\ln\ln n.
\label{dsca}
\eeq
The data for $D(n)$, to be presented in Figure~\ref{idcross},
are indeed observed to grow very slowly with time $n$.
A quantitative check of the asymptotic doubly logarithmic dependence is however
practically impossible.
The lead persistence probability $S(n)$,
i.e., the probability that the first node has kept the lead up to time $n$,
is then expected to fall off exponentially with $D(n)$, i.e., as
\beq
S(n)\sim1/(\ln n)^\theta.
\eeq

\subsection{High-temperature regime: crossover to the Barab\'asi-Albert (BA) model}
\label{highcross}

In the $T\to\infty$ limit, the BB model becomes the BA model.
For any finite time $n$,
every quantity in the BB model goes continuously
to the analogous quantity in the BA model.
Some asymptotic long-time quantities may however exhibit discontinuous behavior.
In technical words, the $n\to\infty$ and $T\to\infty$ limits need not commute.
An example is the asymptotic probability $\Pi_\infty(T)$ for the leader to be a record,
shown in Figure~\ref{q}.
The crossover scale can be determined by considering the index of the leader.
In the BA model, the mean index of the leader has finite asymptotic values
which depend on the initial state~\cite{L},
namely $\mean{I}_\infty^\A\approx3.40$ and $\mean{I}_\infty^\B\approx2.67$.
In the BB model, the prefactor of the scaling law~(\ref{wsca})
vanishes in the $T\to\infty$ limit.
The crossover in the $n$-$T$ plane takes place for a time $n\sim\tau_\BA(T)$, with
\beq
\ln\tau_\BA(T)\sim T\qquad(T\to\infty).
\eeq
For a fixed high temperature $T\gg1$,
the BB model behaves essentially as the BA model for $n\ll\tau_\BA(T)$,
whereas the asymptotic behavior of the BB model only appears for $n\gg\tau_\BA(T)$.

\subsection{Low-temperature regime: crossover to the record-driven (RD) growth process}
\label{lowcross}

In the $T\to0$ limit,
the BB model becomes the record-driven (RD) growth process~\cite{R}.
In the latter model, the current record node attracts all the new connections,
until it is outdone by the next record.
Here again, some asymptotic long-time quantities may exhibit discontinuous behavior.
The crossover scale can again be estimated by considering the index of the leader.
The following line of reasoning holds irrespectively of the presence ($\theta>1$)
or of the absence ($\theta\le1$)
of a condensation transition at a finite critical temperature $T_\c$.
In the RD growth process,
as a consequence of the scale invariance of the underlying record process,
the mean index of the leader grows linearly with time~$n$, as $\mean{I(n)}\approx\chi n$.
The analytic expression of the amplitude $\chi\approx0.275\,765$
is derived in Appendix~B (see~(\ref{sres})), using the formalism of~\cite{R}.
In the BB model, the prefactor of the scaling law~(\ref{wsca}) diverges as $T\to0$.
The crossover time $\tau_\RD(T)$ is such that $((\ln\tau_\RD)/T)^\theta\sim\tau_\RD$,
hence, to leading~order,
\beq
\tau_\RD(T)\sim T^{-\theta}\qquad(T\to0).
\label{xird}
\eeq
For a fixed low temperature $T\ll1$,
the BB model behaves essentially as the RD process for $n\ll\tau_\RD(T)$,
whereas the true asymptotic behavior only appears for $n\gg\tau_\RD(T)$.

\begin{figure}
\begin{center}
\includegraphics[angle=-90,width=.5\linewidth]{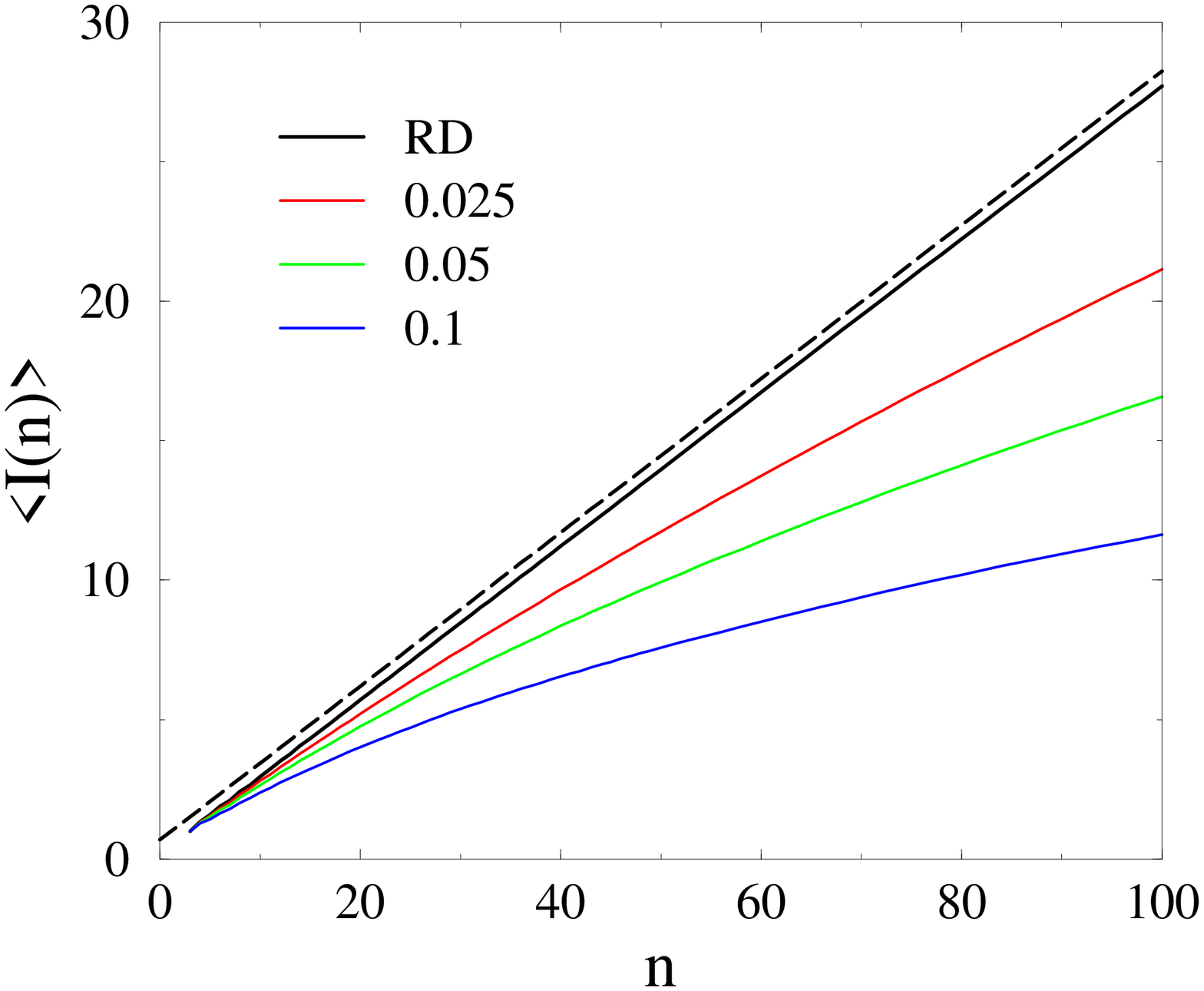}
\includegraphics[angle=-90,width=.48\linewidth]{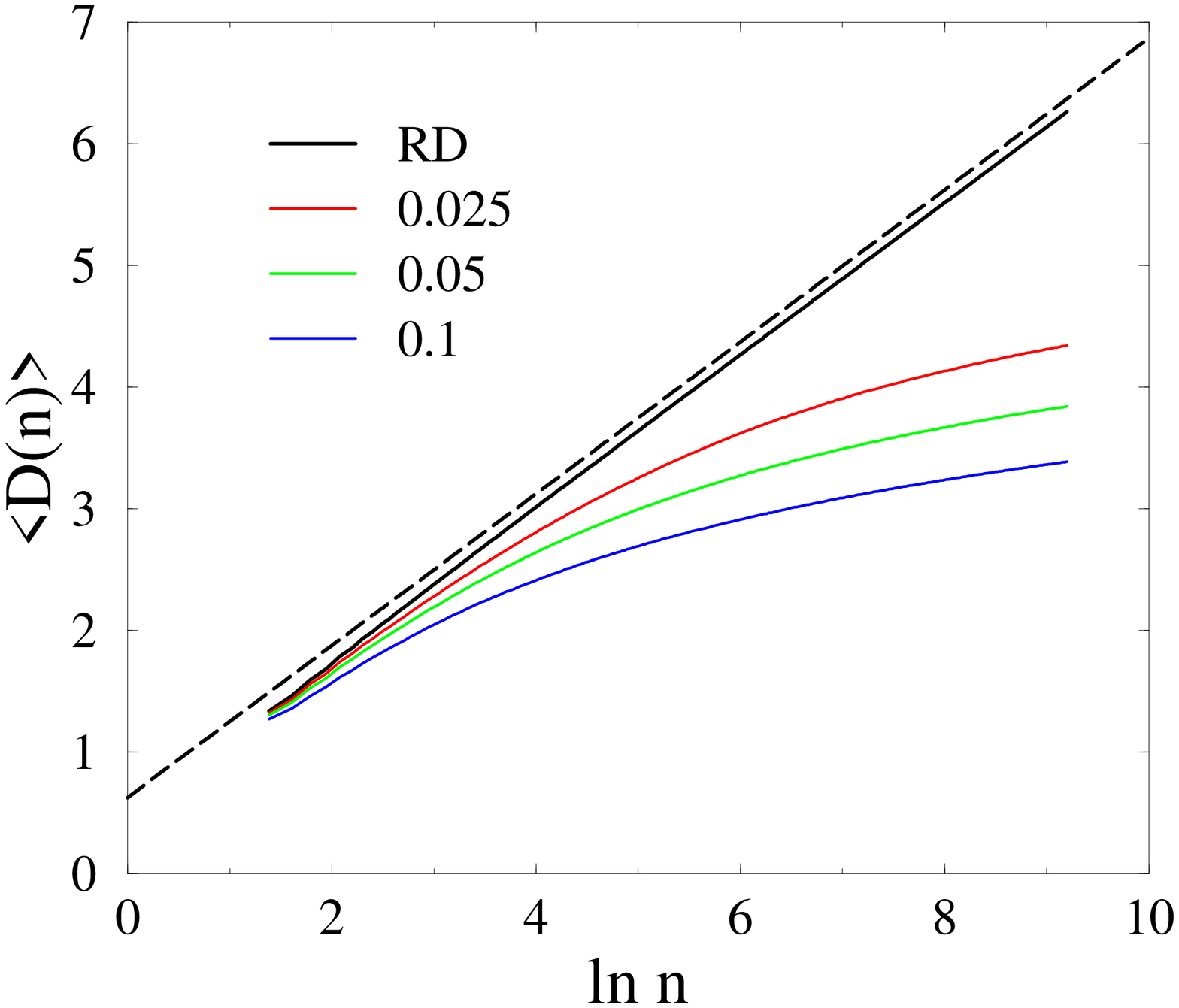}
\caption{\label{idcross}
Data for the BB model for three low values of temperature $T$ and initial condition~B
(colors) and for the RD process (black).
Left: mean index $\mean{I(n)}$ of the leader against $n$.
Right: mean number $\mean{D(n)}$ of distinct leaders against $\ln n$.
The dashed lines have the theoretical asymptotic slopes $\chi$ and $\omega$.}
\end{center}
\end{figure}

Figure~\ref{idcross} gives an illustration of the crossover of the
BB model to the RD model on two characteristic quantities.
The left panel shows $\mean{I(n)}$ against $n$.
The RD data closely follow the linear law with slope $\chi$ (dashed line),
whereas the BB data bend toward the logarithmic growth~(\ref{wsca}).
The right panel shows $\mean{D(n)}$ against $\ln n$.
The RD data closely follow the linear law with slope
$\omega\approx0.624\,330$ (dashed line) (see~(\ref{rres})),
whereas the BB data bend toward the doubly logarithmic growth~(\ref{dsca}).

\begin{figure}
\begin{center}
\includegraphics[angle=-90,width=.5\linewidth]{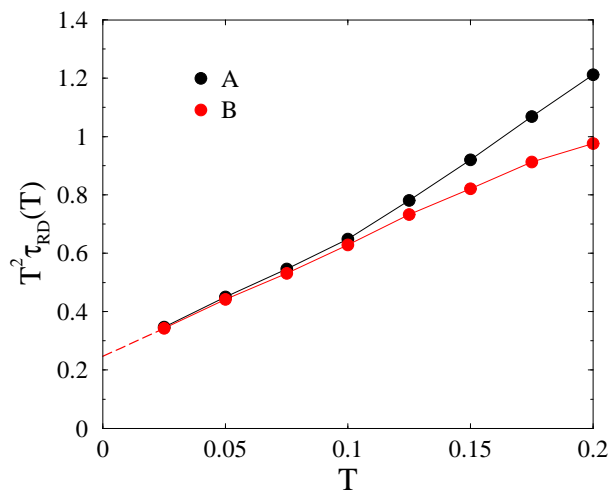}
\caption{\label{tcross}
Plot of $T^2$ times the crossover time $\tau_\RD(T)$
against temperature $T$ for both initial conditions.
Dashed line: extrapolation of a common fit to the data, yielding the intercept 0.25.}
\end{center}
\end{figure}

From an operational viewpoint, for a fixed temperature $T$,
the crossover time can be defined as being the time
for which $\mean{I(n,T)}=\mean{I(n,0)}/2$ is half its value in the RD growth process.
Figure~\ref{tcross} shows $T^2$ times the crossover time $\tau_\RD(T)$
thus defined, against temperature $T$, for both initial conditions.
The measured intercept yields the behavior $\tau_\RD(T)\approx 0.25/T^2$,
in agreement with~(\ref{xird}).

\section{Degree statistics in the condensed phase:
an infinite hierarchy of flu\-c\-tu\-a\-ting condensates}
\label{snapshot}

The goal of this section is to give a picture for the dynamics
of the condensates in the low-temperature phase ($T<T_\c$).
To do so, we investigate the statistics of the ordered sequence of largest degrees.
For a given large time $n$,
we rank the $n$ nodes of the network in order of decreasing degrees:
\beq
k\lab{1}\ge k\lab{2}\ge k\lab{3}\ge\dots,
\eeq
and we set
\beq
k\lab{j}=R\lab{j}n.
\eeq
The reduced variable $R\lab{j}$ represents the `size' or `weight'
of the $j$-th largest degree,
i.e., the degree fraction carried by the corresponding node.
In particular $k\lab{1}=k_\max$ is the degree of the leader at time $n$,
and $R\lab{1}$ is the degree fraction carried by this leader.

Let us recall that, according to the continuum approach of~\cite{bb1,bb2},
the total size of all the condensates
is equal to the following deterministic (i.e., non-fluctuating) quantity:
\beq
F=\sum_jR\lab{j}=1-K(T,1).
\eeq
(see~(\ref{stotal})).
As already discussed in Section~2.1,
the continuum approach does not say anything about the number of condensates,
nor about the mean values
or the fluctuations of the individual sizes $R\lab{j}$ of the condensates.

\begin{figure}
\begin{center}
\includegraphics[angle=-90,width=.48\linewidth]{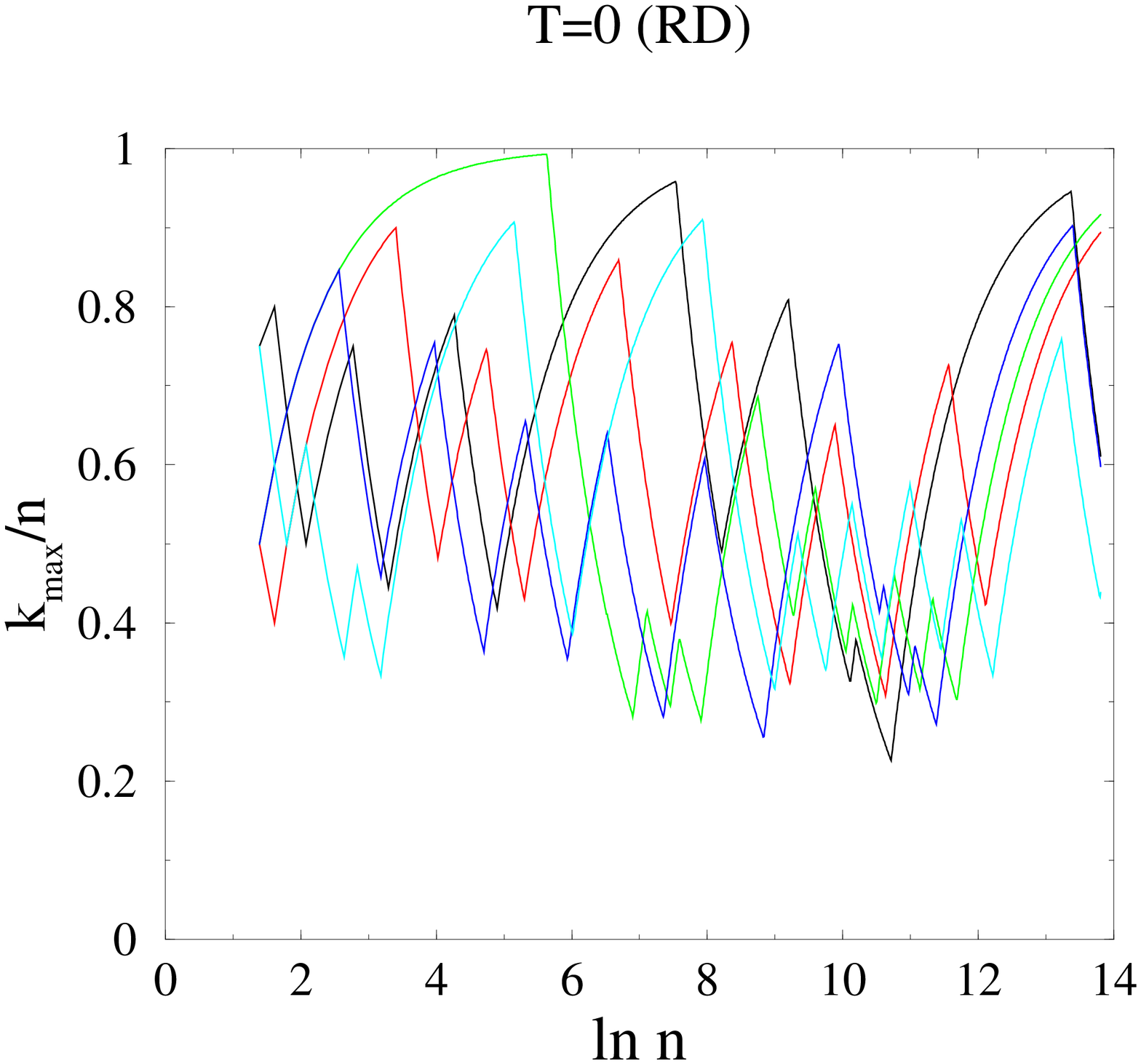}
\includegraphics[angle=-90,width=.48\linewidth]{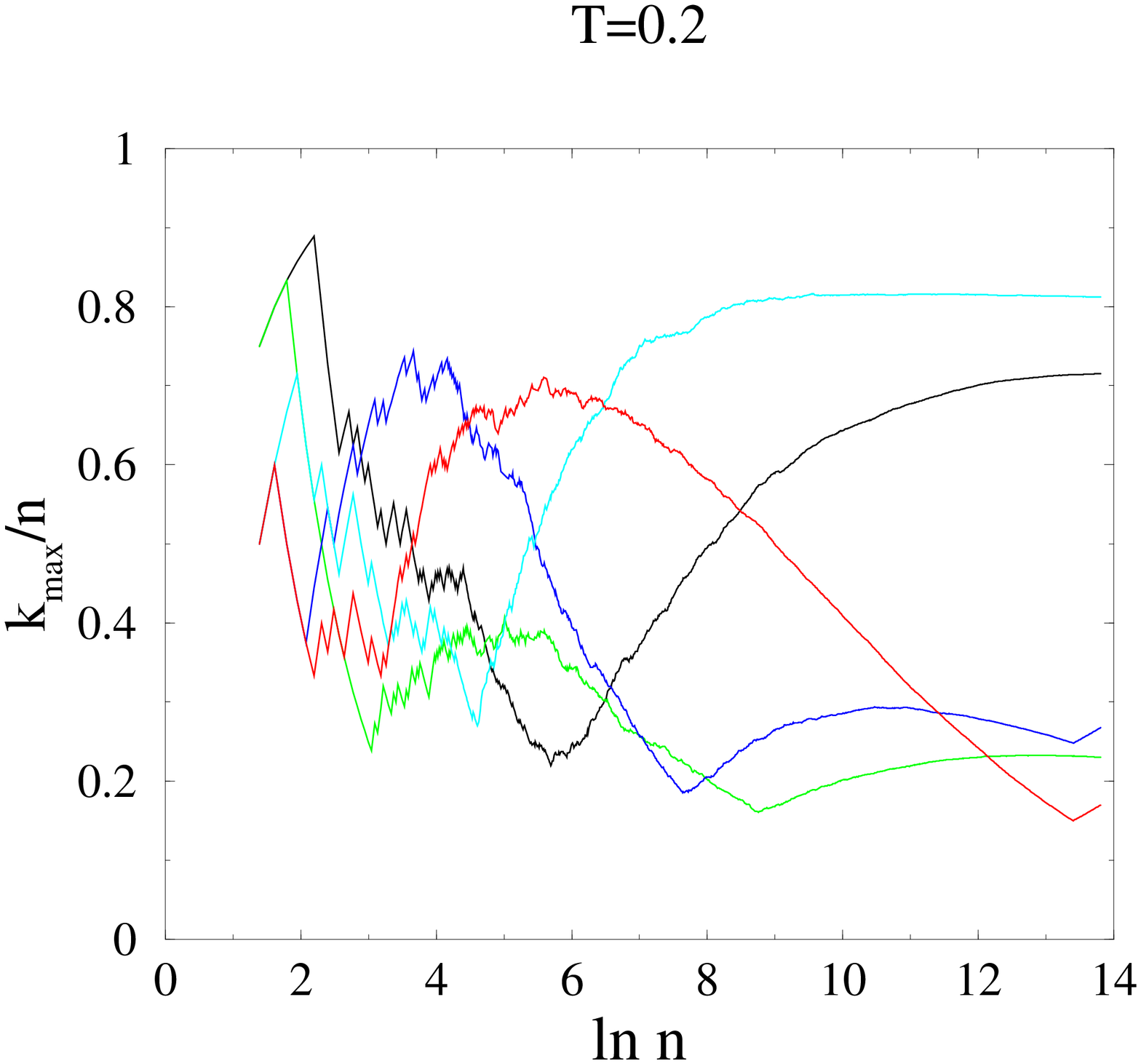}
\caption{\label{kmax}
Five typical tracks $R\lab{1}=k_\max/n$ against $\ln n$ for initial condition~A.
Left: RD growth process (i.e., $T=0$ limit).
Right: $T=0.2$.}
\end{center}
\end{figure}

A first qualitative picture is given in Figure~\ref{kmax},
showing five typical tracks of the degree fraction $R\lab{1}=k_\max/n$ of the leader
against $\ln n$.
The left panel shows data in the RD growth process ($T=0$),
whereas the right one shows data at temperature $T=0.2$.
In both situations the degree of the leader exhibits large fluctuations
around its mean value, that is $\mean{R\lab{1}}=\omega\approx0.624\,330$ (see~(\ref{rres}))
for the RD growth process,
and $\mean{R\lab{1}}\approx0.46$ for $T=0.2$.
In the RD growth process, because of the scale invariance of the underlying record process,
we observe a stationary signal on a logarithmic scale.
At finite temperature, the tracks first resemble those of the RD process,
however with some noise on a short time scale.
The dynamics then slows down drastically when time
exceeds a few times the crossover time $\tau_\RD$ (see~(\ref{xird})).
We have $\ln\tau_\RD\approx3.41$ for $T=0.2$ and initial condition A,
whereas the rather sudden slowing down is observed near $\ln n\approx4.5$.
Let us however note that fluctuations in the degree of the leader
do persist forever, on a very slow scale, though.
The number of maxima or minima is indeed expected to grow
proportionally to the number of distinct leaders,
i.e., as $\ln\ln n$ (see~(\ref{dsca})).

\begin{figure}
\begin{center}
\includegraphics[angle=-90,width=.45\linewidth]{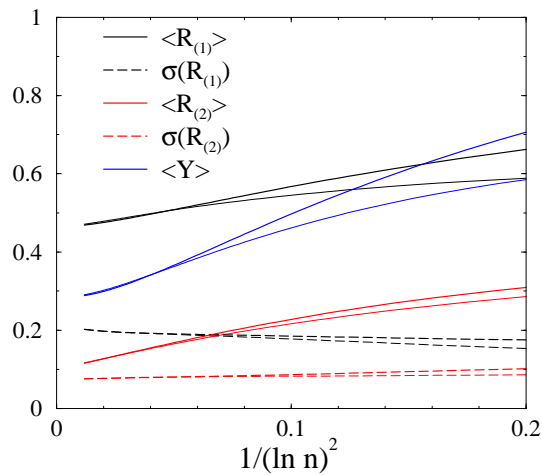}
\caption{\label{rtest}
Mean values and standard deviations
of the sizes of the larger two `condensates' $(j=1,2)$, and mean value $\mean{Y}$,
at temperature $T=0.2$, against $1/(\ln n)^2$.
Thin (resp.~thick) lines show data for initial condition~A (resp.~B).}
\end{center}
\end{figure}

Turning to a more quantitative viewpoint,
we present in Figure~\ref{rtest} the mean values $\mean{R\lab{j}}$
and standard deviations $\sigma(R\lab{j})=(\mean{R\lab{j}^2}-\mean{R\lab{j}}^2)^{1/2}$
of the larger two condensates $(j=1,2)$, again at temperature $T=0.2$.
The mean value $\mean{Y}$ of the moment variable $Y$, to be defined below
(see~(\ref{ydef})), is also plotted.
The data have the following limiting values, irrespectively of the initial condition:
$\mean{R\lab{1}}\approx0.461$, $\sigma(R\lab{1})\approx0.213$,
$\mean{R\lab{2}}\approx0.100$, $\sigma(R\lab{2})\approx0.075$,
and $\mean{Y}\approx0.280$.
The choice $1/(\ln n)^2$ of the abscissa is motivated theoretically
by the prediction~(\ref{kcsca}) for the largest degree at the critical point.
A phenomenological analysis of the data, e.g.~by fitting a power-law fall-off,
would essentially yield the same asymptotic values.

Several conclusions can be drawn from the above results.
The extrapolated values of $\mean{R\lab{1}}$ and $\mean{R\lab{2}}$
are clearly different from zero,
indicating that the condensed phase of the BB model
is characterized by the presence of more than one condensate.
The widths $\sigma(R\lab{j})$ are comparable to the means $\mean{R\lab{j}}$,
demonstrating that the sizes $R\lab{j}$ are non-self-averaging.

The above observations point toward the scenario of an infinite
hierarchically ordered sequence of fluctuating condensates.
This situation is thus qualitatively similar to that of the RD growth process
describing the model at zero temperature, recalled in Appendix~B.
Namely, the reduced variables $R\lab{j}$ ($j=1,2,\dots$)
have a non-trivial joint distribution
in the limit of a large network $(n\to\infty)$.
In other words, the condensed phase is characterized by the presence
of an infinite hierarchy of condensates,
whose sizes $R\lab{j}$ are non-self-averaging but keep fluctuating forever.
The mean sizes of the successive condensates however fall off rapidly
with the rank index $j$.

One convenient tool to investigate this collection of fluctuating condensates
consists in introducing the second-moment variable
\beq
Y(n)=\sum_{i=1}^n\left(\frac{k_i(n)}{n}\right)^2
\label{ydef}
\eeq
which converges to the asymptotic variable
\beq
Y=\sum_{j\ge1}R\lab{j}^2.
\label{yasy}
\eeq
The second-moment variable $Y$ (and its higher-order generalizations $Y^p$)
have been introduced seemingly for the first time by Derrida and Flyvbjerg~\cite{df},
in an investigation of the sequence of weights obtained by the iterative random breaking
of an interval.
The data plotted in Figure~\ref{rtest} (blue lines) suggest the existence of
such a fluctuating limiting variable $Y$, such that $\mean{Y}\approx0.280$ at $T=0.2$.

\begin{figure}
\begin{center}
\includegraphics[angle=-90,width=.45\linewidth]{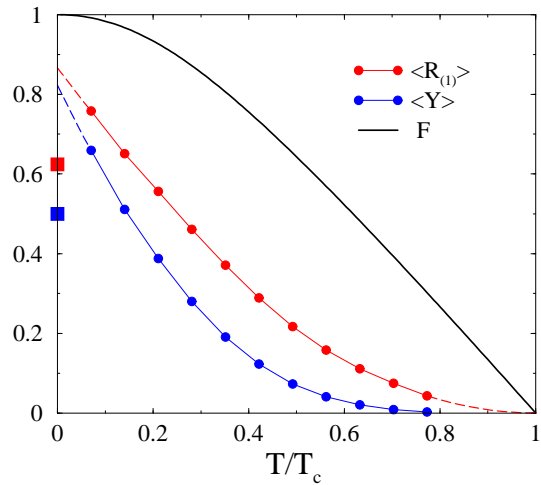}
\caption{\label{cond}
Black line: theoretically predicted condensed fraction
$F$ against reduced temperature $T/T_\c$.
Symbols: numerical data for the mean degree fraction
of the leader $\mean{R\lab{1}}$ (red)
and the mean moment variable $\mean{Y}$ (blue).
Symbols on left axis: corresponding quantities in the RD process.}
\end{center}
\end{figure}

The dependence of the distribution of the hierarchy of condensates on temperature
is illustrated in Figure~\ref{cond}.
The theoretically predicted total condensed fraction $F$
(see~(\ref{stotal})) is plotted against reduced temperature $T/T_\c$ (thick black line),
as well as data for the mean degree fraction of the leader $\mean{R\lab{1}}$ (red symbols)
and for the mean moment variable $\mean{Y}$ (blue symbols).
All over the condensed phase,
the mean degree fraction $\mean{R\lab{1}}$ of the leader
is observed to be significantly smaller than the total condensed fraction.
This demonstrates that the multiplicity of condensates is a significant phenomenon
over the whole condensed phase.

\begin{figure}
\begin{center}
\includegraphics[angle=-90,width=.5\linewidth]{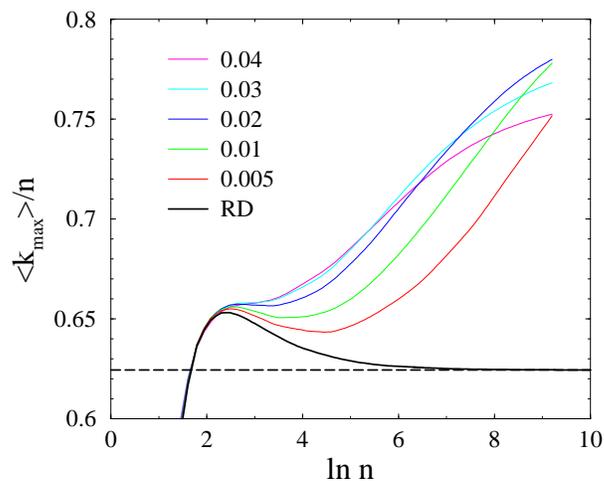}
\caption{\label{kcross}
Mean degree fraction of the leader $\mean{R\lab{1}}=\mean{k_\max}/n$
against $\ln n$, for the BB model with initial condition~A
at several low temperatures (colors) and for the RD process (black).
Dashed line: limiting value $\omega$ for the RD process.}
\end{center}
\end{figure}

We now turn to a more detailed study of the two endpoints of the condensed phase,
corresponding respectively to $T\to0$ and $T\to T_\c$.
The low-temperature ($T\to0$) regime exhibits a crossover to the RD growth process,
as already underlined in Section~\ref{lowcross}.
The large symbols on the vertical axis of Figure~\ref{cond}
show the corresponding values in the RD growth process,
namely $\mean{R}=\omega\approx0.624\,330$ (see~(\ref{rres}))
and $\mean{Y}=1/2$ (see~(\ref{yres})).
These numbers are clearly far below the extrapolated $T\to0$ limits
$\mean{R\lab{1}}\approx0.865$ and $\mean{Y}\approx0.82$,
thus providing explicit examples
of the discontinuous behavior announced in Section~\ref{lowcross}.
The manifestation of the RD to BB crossover on the degree of the leader
is illustrated in Figure~\ref{kcross}.
For a fixed low temperature $T$,
the mean degree fraction of the leader, $\mean{R\lab{1}}=\mean{k_\max}/n$,
first follows the behavior characteristic of the RD process,
up to a time of order the crossover time $\tau_\RD(T)$ (see~(\ref{xird})),
tending to converge to the RD limiting value $\omega$ (dashed line),
before it bends away from $\omega$ and eventually saturates to a much higher value,
close to the $T\to0$ limit $\mean{R\lab{1}}\approx0.865$.

In the vicinity of the critical temperature ($T\to T_\c^-$),
the total condensed fraction $F\approx\mu(T_\c)\Delta T$
vanishes linearly with $\Delta T=T_\c-T$.
The corresponding slope $\mu(T_\c)$ has been evaluated in~(\ref{tc}).
The mean degree fraction of the leader $\mean{R\lab{1}}$ is observed
to fall off much faster than linearly as $\Delta T\to0$,
thus indicating that the leader carries a smaller and smaller part
of the total size of the condensates as the critical point is approached.
This observation can be corroborated by means of the following line of reasoning.
Right at the critical point ($T=T_\c$),
the degree of the leader is predicted to grow as~(\ref{kcsca}).
The corresponding degree fraction therefore vanishes logarithmically slowly,
as $R\lab{1}\sim1/(\ln n)^\theta$.
Now, with the natural assumption that the degree of the leader obeys
a finite-size scaling law around the critical point,
the above time dependence at $T=T_\c$ can be turned, using~(\ref{dtc}), to the estimate
\beq
R\lab{1}\sim(\Delta T)^{\theta/(\theta-1)}.
\label{rc}
\eeq
For $\theta=2$ we obtain the quadratic behavior $R\lab{1}\sim(\Delta T)^2$,
in agreement with the data shown in Figure~\ref{cond}.

The scaling law~(\ref{rc}) can be rephrased in the following form.
The effective number of condensates,
defined in an operational way as the ratio $J=F/\mean{R\lab{1}}$,
diverges as
\beq
J\sim(\Delta T)^{-1/(\theta-1)}
\label{jc}
\eeq
in the vicinity of the critical temperature ($T\to T_\c^-$).
The exponents of the power laws~(\ref{rc}) and~(\ref{jc})
diverge as $\theta\to1$, where the condensation transition disappears.

\section{Discussion}

After~\cite{R,L,D},
the present work closes up a cycle of four papers
devoted to the degree statistics in growing networks with preferential attachment,
the main emphasis being on leaders and lead changes.
We have focussed most of our attention onto
the Bianconi-Barab\'asi (BB) fitness model~\cite{bb1,bb2}
and on its infinite-temperature and zero-temperature limits,
respectively corresponding to
the Bianconi-Barab\'asi (BA) model~\cite{ba}
and the record-driven (RD) growth model~\cite{R}.
Very few works had been devoted to leaders in growing structures
so far~\cite{el,lkr,lm,lbk}.

Our main findings concerning the history of leaders can be summed up as follows.

\noindent {\bf (i)} At infinite temperature (BA model),
the leader statistics keeps a memory of the initial state of the network.
A typical history of the network involves finitely many distinct leaders,
chosen among the oldest nodes~\cite{L}.
The mean index of the leader, the mean number of distinct leaders,
and the lead survival probability
have finite asymptotic values, which depend on the initial condition.

\noindent {\bf (ii)} At zero temperature (RD process),
the history of leaders inherits the temporal self-similarity
of the underlying record process~\cite{R}.
This property manifests itself as stationarity on a logarithmic scale.
There are typically $\ln n$ records and $\omega\,\ln n$ distinct leaders up to time $n$.
The degree and the index of the leader grow linearly with time,
as $k_\max\approx Rn$ and $I\approx Sn$,
where $R$ and $S$ keep fluctuating,
and admit non-trivial limiting distributions so that $\mean{R}=\omega\approx0.624\,330$, and $\mean{S}=\chi\approx0.275\,765$.

\noindent {\bf (iii)} In the generic finite-temperature situation,
the main features of the history of leaders hold irrespectively of temperature,
i.e., both in the fluid phase ($T>T_\c$) and in the condensed phase ($T<T_\c$).
The leader is a record node with high probability,
whereas its index grows on a logarithmic scale, as $I\sim((\ln n)/T)^\theta$.
The number of distinct leaders up to time~$n$ therefore grows forever,
although extremely slowly, on a doubly logarithmic scale: $D(n)\sim\ln\ln n$.
This picture can be captured into the motto:
{\it The Leader of Today is the Record of Old.}

The above scenario for the history of leaders relies on one essential ingredient,
which was missing in the work by Ferretti and Bianconi~\cite{fb},
namely the separation of time scales induced by the very slow growth
of the index of the leader.
This explains why the conclusions of the present work are in many respects
more precise than those of~\cite{fb}.

When temperature is either very high, or very low, the generic behavior
of the BB model is eventually observed beyond a crossover time scale
$\tau_\BA(T)$ which diverges at high temperature
(whereas BA behavior is observed for shorter times)
or $\tau_\RD(T)$ which diverges at low temperature
(whereas RD behavior is observed for shorter times).
As a consequence, histories ending with a single final leader
never occur except in the BA model, i.e., strictly at infinite temperature.

\begin{figure}
\begin{center}
\includegraphics[angle=-90,width=.35\linewidth]{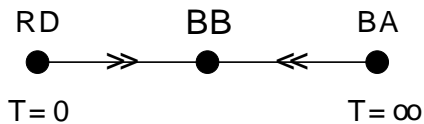}
\caption{\label{rg}
Schematic renormalization-group flow of the model.}
\end{center}
\end{figure}

The above discussion can be recast in the language of the renormalization group.
The BA and RD models respectively appear
as infinite-temperature and zero-temperature unstable fixed points,
whereas the BB model corresponds to a stable finite-temperature fixed point.
The associated renormalization-group flow is sketched in Figure~\ref{rg}.
The situation is the exact opposite of that usually met in critical phenomena.
Furthermore, the fixed point describing the critical behavior around
the condensation transition temperature $T_\c$ is not visible in this flow.
Table~\ref{summary} summarizes the asymptotic behavior of three key quantities
(number of distinct leaders, index of the leader, degree of the leader)
in the various regimes of the model.

\begin{table}
\begin{center}
\begin{tabular}{|c|c|c|c|}
\hline
Quantity&BA ($T=\infty$)&BB ($T$ finite)&RD ($T=0$)\\
\hline
$\mean{D(n)}$&$\mean{D}_\infty$ finite&$\theta\,\ln\ln n$&$\omega\,\ln n$\\
\hline
$\mean{I(n)}$&$\mean{I}_\infty$ finite&$((\ln n)/T)^\theta$&$\chi n$\\
\hline
\treshaut
$\mean{k_\max(n)}$&$n^{1/2}$
&$\matrix{
T>T_\c:\hfill&n^{1/C(T)}\hfill\cr
T=T_\c:\hfill&n/(\ln n)^\theta\hfill\cr
T<T_\c:\hfill&\mean{R\lab{1}}n\hfill
}$
&$\omega n$\\
\hline
\end{tabular}
\end{center}
\caption{Asymptotic temporal behavior of the key quantities:
mean number $\mean{D(n)}$ of distinct leaders,
mean index $\mean{I(n)}$ of the leader,
mean degree $\mean{k_\max(n)}$ of the leader, in the various regimes of the model.
Only the last observable is sensitive to the presence of a condensation transition
at a finite temperature $T_\c$.}
\label{summary}
\end{table}

All the predictions recalled so far are insensitive to the presence
of a condensation transition at a finite temperature $T_\c$.
Only the last observable $\mean{k_\max(n)}$ depends on the position of temperature $T$
with respect to $T_\c$.
If the BB model has no condensation transition ($\theta\le1$),
the subextensive growth law $\mean{k_\max(n)}\sim n^{1/C(T)}$
holds for all positive temperatures.

Another motivation for the present work was the desire of a better understanding
of the dynamics in the condensed phase ($T<T_\c$).
On the one hand,
the analogy with the Bose-Einstein mechanism put forward originally in~\cite{bb2}
suggests a single static condensate with degree $k_\max(n)\approx Fn$,
whose relative size $F$ is predicted by the mean-field-like approach
(see~(\ref{stotal})),
and therefore does not fluctuate in time.
On the other hand, for some other stochastic processes leading to condensation,
such as the zero-range process,
the condensate is known to perform some ergodic stationary-state motion over the system,
on a time scale which diverges as a power of the system size (spatial extent)~\cite{uszrp}.
The situation of growing structures is made more complex by the fact that
the system size is time itself, so that strictly speaking there is no stationary state.

At variance with earlier studies~\cite{bb2,fb},
the present work leads to the picture of an infinite hierarchy of condensates,
whose degrees grow linearly with time, as $k\lab{j}\approx R\lab{j}n$.
The relative sizes $R\lab{j}$,
generalizing the reduced degree of the leader $R\lab{1}=R=k_\max/n$,
sum up to the predicted static condensed fraction $F$.
The individual sizes $R\lab{j}$ of the condensates are however
non-self-averaging.
They rather keep fluctuating forever,
and admit non-trivial temperature-dependent limiting distributions
over the whole low-temperature condensed phase.

\subsection*{Acknowledgments}

It is a pleasure to thank G Bianconi for having raised our interest
in the nature of the condensate in the low-temperature phase of the BB model.

\appendix
\section{Extreme-value statistics for i.n.i.d.~variables: an example}

In this Appendix we investigate the sequence of i.n.i.d.~random variables:
\beq
Z_i=i\,U_i\qquad(i=1,\dots,n),
\label{iui}
\eeq
where the $U_i$ are i.i.d.~variables uniformly distributed on the interval $[0,1]$.
We are mostly interested in the distribution
of the largest value $Z_\max$ among the $n$ random numbers $Z_i$,
and of the index $\Iinid$ where this largest value is reached.
Our goal is to test the variational approach proposed in Section~\ref{variational},
by comparing its predictions to the exact results derived below.

\subsection*{A.1.~Variational approach}

In the present situation, the variational approach proceeds as follows.
The random variable~$Z_i$ lies in the interval $[0,i]$,
whose upper bound $i$ takes its largest value for $i=n$.
The index $\Iinid$ is therefore expected to be localized
in a relatively small range of width $w$ around~$n$, i.e., $n-\Iinid\sim w$.
On the other hand,
the largest value $U_\max$ among the $w$ variables~$U_i$ sampled in that range
is such that the probability of having $U>U_\max$,
i.e., $1-U_\max$, is of order $1/w$.
Neglecting every numerical prefactor,
the estimates $n-\Iinid\sim w$ and $1-U_\max\sim1/w$ yield
\beq
n-Z_\max\sim w+\frac{n}{w}.
\eeq
This expression has a non-trivial minimum for $w=\sqrt{n}$, where both terms are equal.
The variational approach thus predicts the following scaling relations for large $n$:
\beq
n-Z_\max\approx2(n-\Iinid)\sim\sqrt{n}.
\label{varres}
\eeq

\subsection*{A.2.~Exact results for finite $n$}

The example~(\ref{iui}) is simple enough
to allow one to derive exact results on the distribution of $Z_\max$ and $\Iinid$,
for a finite number $n$ of variables.

Let us start with the largest value $Z_\max$.
The common distribution function of the uniform variables $U_i$ on $[0,1]$ reads
\beq
F_U(u)=\prob{U_i<u}=\left\{\matrix{
0\hfill&\hbox{for}&u\le 0,\hfill\cr
u\hfill&\hbox{for}&0\le u\le1,\hfill\cr
1\hfill&\hbox{for}&u\ge1.\hfill}\right.
\eeq
The distribution function of $Z_\max$ is therefore
\beq
F_{Z_\max}(z)=\prob{Z_\max<z}=\prod_{i=1}^n\prob{iU_i<z}
=\prod_{i=1}^nF_U\left(\frac{z}{i}\right).
\eeq
Setting $\Int z=n-k$, the latter expression reads explicitly
\beq
F_{Z_\max}(z)=\prod_{i=n-k+1}^n\frac{z}{i}=\frac{(n-k)!\,z^k}{n!}.
\label{fz}
\eeq
The mean value of $Z_\max$ is given by
\beq
\mean{Z_\max}=\int_0^nF'_{Z_\max}(z)\,z\,\d z=\int_0^n(1-F_{Z_\max}(z))\,\d z.
\eeq
Using the expression~(\ref{fz}), we obtain
\beqa
\mean{Z_\max}&=&n-\sum_{k=1}^n\frac{(n-k)!}{n!}\int_{n-k}^{n-k+1}z^k\,\d z
\nonumber\\
&=&n-\sum_{k=1}^n\frac{(n-k)!}{n!}\,\frac{(n-k+1)^{k+1}-(n-k)^{k+1}}{k+1}
\nonumber\\
&=&\sum_{k=0}^n\frac{(n-k)!}{n!}\,\frac{(n-k)^{k+1}}{(k+1)(k+2)}.
\label{zexact}
\eeqa

Let us now turn to the index $\Iinid$.
Consider a fixed index $i=1,\dots,n$ and a fixed value $z_i$ of the random variable $Z_i$.
For any $j\ne i$, the probability to have $Z_j<z_i$, i.e., $U_j<z_i/j$, reads $F_U(z_i/j)$.
The conditional probability that $Z_i$ is the largest therefore reads
\beq
\prob{Z_i=Z_\max\vert i,z_i}=\prod_{j\ne i}F_U\left(\frac{z_i}{j}\right).
\eeq
Setting $\Int z_i=n-k$, we have explicitly
\beq
\prob{Z_i=Z_\max\vert i,z_i}=\frac{(n-k)!}{n!}\,i\,z_i^{k-1}.
\eeq
The distribution $P_i=\prob{\Iinid=i}$ of the index $\Iinid$
can be obtained by integrating the latter expression against the distribution of $Z_i$:
\beqa
P_i&=&\frac{1}{i}\int_0^i\prob{Z_i=Z_\max\vert i,z_i}\,\d z_i
\nonumber\\
&=&\sum_{k=n-i+1}^n\frac{(n-k)!}{n!}\int_{n-k}^{n-k+1}z_i^{k-1}\,\d z_i
\nonumber\\
&=&\sum_{k=n-i+1}^n\frac{(n-k)!}{n!}\,\frac{(n-k+1)^k-(n-k)^k}{k}.
\label{pi}
\eeqa
The mean value of the index $\Iinid$ is given by
\beqa
\mean{\Iinid}&=&\sum_{i=1}^n i\,P_i
\nonumber\\
&=&\frac{1}{2}\sum_{k=1}^n\frac{(n-k)!}{n!}\,\left((n-k+1)^k-(n-k)^k\right)(2n+1-k).
\nonumber\\
&=&n-\frac{1}{2}\sum_{k=1}^n\frac{(n-k)!}{n!}\,(n-k)^k.
\label{iexact}
\eeqa
Table~\ref{zitable} gives the first few values of $\mean{Z_\max}$ and $\mean{\Iinid}$,
as given by the exact rational expressions~(\ref{zexact}) and~(\ref{iexact}).
In spite of the simplicity of these numbers,
which calls for a combinatorial interpretation,
we have not been able to relate them to anything known in the literature.

\begin{table}
\begin{center}
\begin{tabular}{|c|c|c|c|c|c|c|c|}
\hline
$n$&1&2&3&4&5&6&7\\
\hline
\haut $\mean{Z_\max}$&$\frad{1}{2}$&$\frad{13}{12}$&$\frad{125}{72}$&
$\frad{3503}{1440}$&$\frad{5687}{1800}$&$\frad{49253}{12600}$
&$\frad{19797847}{4233600}$\\
\hline
\haut $\mean{\Iinid}$&1&$\frad{7}{4}$&$\frad{31}{12}$
&$\frad{55}{16}$&$\frad{1033}{240}$&$\frad{829}{160}$
&$\frad{61153}{10080}$\\
\hline
\end{tabular}
\end{center}
\caption{Mean largest value $\mean{Z_\max}$ and mean index $\mean{\Iinid}$,
as given by the exact results~(\ref{zexact}) and~(\ref{iexact}), for $n$ up to 7.}
\label{zitable}
\end{table}

\subsection*{A.3.~Asymptotic results for large $n$}

Let us now turn to the analysis of the large $n$ behavior of the above results.
The expression~(\ref{fz}) for the distribution function of $Z_\max$
can be estimated as
\beq
F_{Z_\max}(z)\approx\frac{(n-k)!(n-k)^k}{n!}\approx\e^{-k^2/(2n)},
\label{fzasy}
\eeq
where we have used the approximation $z\approx\Int z=n-k$, as well as Stirling's formula.
The above result shows that $n-Z_\max$ scales as $k\sim\sqrt{n}$.
More precisely, setting
\beq
Z_\max=n-X\sqrt{n},
\eeq
and differentiating~(\ref{fzasy}),
we obtain the following distribution for the scaling variable~$X$:
\beq
f_X(x)=x\,\e^{-x^2/2}.
\eeq
Along the same lines, the expression~(\ref{pi}) for the distribution of $\Iinid$
can be recast as
\beq
P_i\approx\frac{1}{n}\int_{n-i}^n\e^{-k^2/(2n)}\,\d k.
\label{piasy}
\eeq
This result shows that $n-\Iinid$ again scales as $k\sim\sqrt{n}$.
Setting
\beq
\Iinid=n-Y\sqrt{n},
\eeq
the expression~(\ref{piasy}) yields the distribution of $Y$:
\beq
f_Y(y)=\int_y^\infty\e^{-x^2/2}\,\d x=\sqrt{\frac{\pi}{2}}\erfc\frac{y}{\sqrt{2}}.
\eeq
The mean values of the scaling variables read
$\mean{X}=\sqrt{\pi/2}$ and $\mean{Y}=\sqrt{\pi/8}$, so that
\beq
\mean{n-Z_\max}\approx2\mean{n-\Iinid}\approx\sqrt{\frac{\pi n}{2}}.
\eeq
This result fully corroborates the outcome~(\ref{varres}) of the variational approach.
The predicted factor 2
exactly holds for the mean values, as we have $\mean{X}=2\mean{Y}$.
More generally, higher moments of the scaling variables
are related through $\mean{X^p}=(p+1)\mean{Y^p}$.

\begin{figure}
\begin{center}
\includegraphics[angle=-90,width=.45\linewidth]{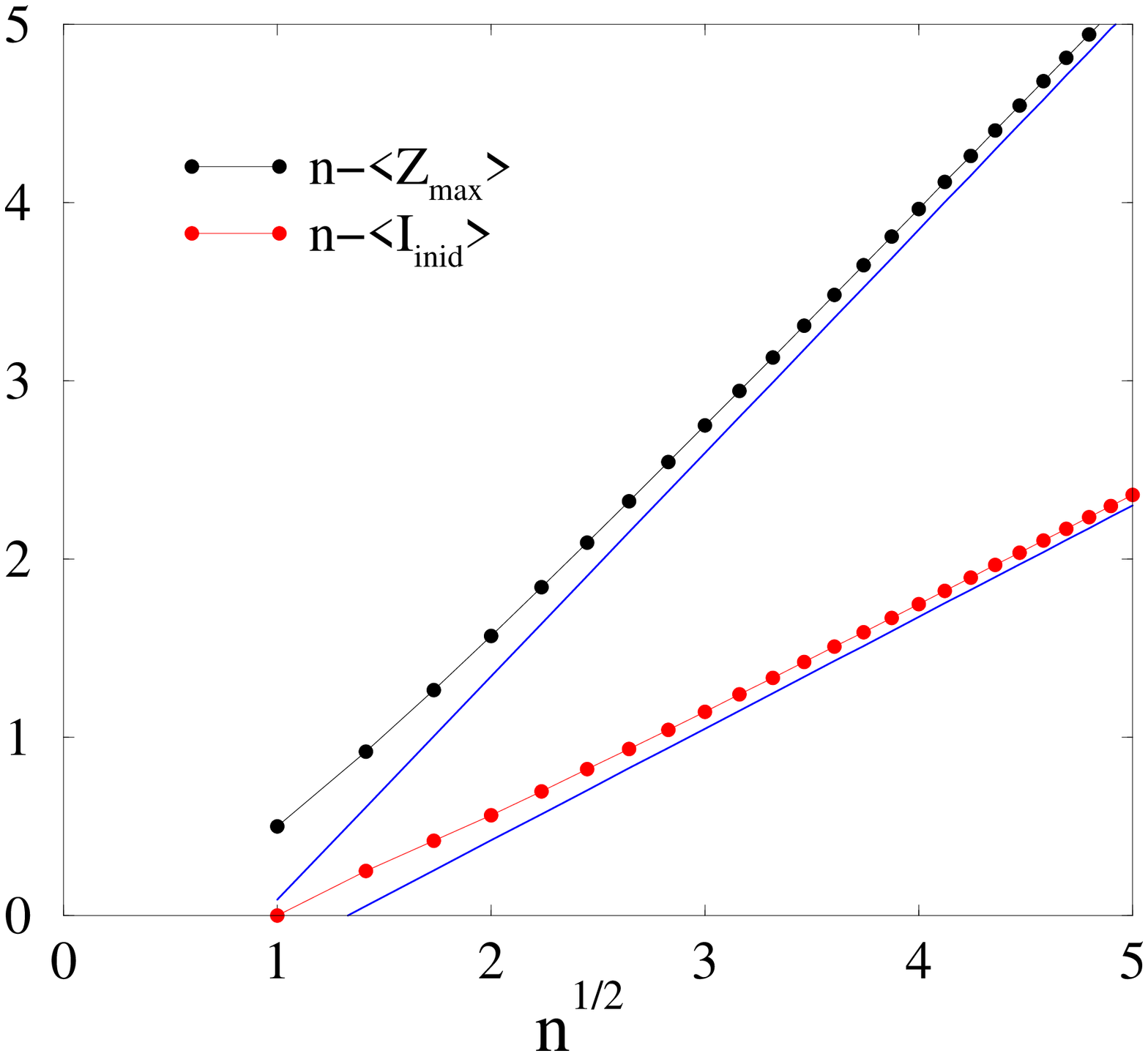}
\caption{\label{zi}
Plot of the differences $n-\mean{Z_\max}$ and $n-\mean{\Iinid}$ against $\sqrt{n}$.
Straight blue lines: asymptotic expansions~(\ref{ziexp}).}
\end{center}
\end{figure}

Finally, corrections to the scaling formulas
for the distributions of $Z_\max$ and $\Iinid$
can also be extracted by a more careful analysis of~(\ref{zexact}) and~(\ref{iexact}).
Skipping details, we just give the expansions for the mean values,
including subleading constant terms:
\beq
\mean{Z_\max}\approx n-\sqrt{\frac{\pi n}{2}}+\frac{7}{6},\qquad
\mean{\Iinid}\approx n-\sqrt{\frac{\pi n}{8}}+\frac{5}{6}.
\label{ziexp}
\eeq
Figure~\ref{zi} shows the differences
$n-\mean{Z_\max}$ and $n-\mean{\Iinid}$ against $\sqrt{n}$.
The straight blue lines show the asymptotic expansions~(\ref{ziexp}).

\section{Some complements on the record-driven growth process}

In this Appendix we give a few useful complements on the record-driven (RD) growth process~\cite{R}.
In the RD process,
only the nodes which belong to the series of records grow by attracting new connections.
We label the record nodes by their rank $m=1,2,\dots$ in the series of records.
We use the asymptotically exact description of the theory of records
in terms of continuous variables.
The $m$-th record node is born at time $N_m$ ($m=1,2,\dots$),
with $N_1=1$, and with the recursion
\beq
N_m=\frac{N_{m-1}}{U_m},
\eeq
where the $U_m$ are uniform i.i.d.~random variables between 0 and 1.

Throughout the following
it will be sufficient to consider the stroboscoped process at record times $n=N_m$.
The distribution of all dimensionless quantities of interest
can indeed be shown~\cite{R}
to be the same at a record time $N_m$ and at a generic time $n$.

\subsection*{B.1.~Distribution of the degree and of the index of the leader}

The leader at time $n=N_m$ is one of the earlier record nodes,
i.e., node $N_\mu$ for some $\mu=1,\dots,m-1$.
In the language of~\cite{R}, $\mu$ is a leader, whereas the subsequent ones
($\mu+1,\dots,m-1$, if any) are non-leaders.
We have therefore
\beq
n=N_m,\qquad I(n)=N_\mu,\qquad k_\max(n)=N_{\mu+1}-N_\mu.
\eeq
The key point of the analysis consists in introducing the reduced variables
\beq
R_m=\frac{k_\max(n)}{n},\qquad
S_m=\frac{I(n)}{n},
\eeq
and in observing that they obey the recursion relations
\beq
\matrix{
0<U_{m+1}<\frad{1}{1+R_m}\hfill&\Longrightarrow&
\left\{\matrix{R_{m+1}=1-U_{m+1},\hfill\cr S_{m+1}=U_{m+1},\hfill}\right.%
\hfill\cr\cr
\frad{1}{1+R_m}<U_{m+1}<1\hfill&\Longrightarrow&
\left\{\matrix{R_{m+1}=U_{m+1}R_m.\hfill\cr S_{m+1}=U_{m+1}S_m.\hfill}\right.
\hfill}
\label{rsrec}
\eeq
These equations can be viewed as defining a stochastic dynamical system
with co-ordinates $R_m$ and $S_m$ subjected to a time-dependent noise $U_m$.
The dynamical system has a unique invariant measure.
It therefore makes sense to consider the limiting variables $R$ and $S$,
and especially their mean values $\mean{R}=\omega$ and $\mean{S}=\chi$, so that
\beq
\mean{k_\max(n)}\approx\omega n,\qquad\mean{I(n)}\approx\chi n.
\eeq

The variables $R_m$ are autonomous, in the sense that
the recursion for the $R_m$ only involves themselves and the noise $U_m$.
The variables $R_m$ are asymptotically distributed according to
the distribution of a variable $R$ with density $f_R$ between 0 and 1.
The Laplace transform of the density $f_R$ reads~\cite{R}:
\beq
\w f_R(s)=\mean{\e^{-s/R}}=1-\e^{-E(s)},
\eeq
with
\beq
E(s)=\int_s^\infty\frac{\e^{-u}\,\d u}{u}.
\eeq
In particular the mean value of $R$ reads $\mean{R}=\omega$, with
\beq
\omega=\int_0^\infty\e^{-s-E(s)}\d s\approx0.624\,329\,988.
\label{rres}
\eeq
This number is known as the Golomb-Dickman constant.
It appeared in the framework of the decomposition
of an integer into its prime factors~\cite{dickman},
and in the study of the longest cycle
in a random permutation~\cite{golomb,goncharov,shepp}.

At variance with the $R_m$, the variables $S_m$ are not autonomous.
The recursion~(\ref{rsrec}) for the $S_m$ indeed involves,
besides themselves and the noise $U_m$,
the variables $R_m$ as well, through the conditioning inequalities.
The study of the variables $S_m$ is therefore more intricate.
In the body of this paper we are mostly interested
in the mean value $\chi=\mean{S}$.
This quantity can be evaluated by means of the formalism of~\cite{R}.
Hereafter we only give the main lines of the derivation, skipping details.
We translate labels of the record nodes by $\mu$, so that record node number 0 is leader,
whereas the subsequent ones, labeled $k=1,\dots$, are non-leaders.
This is precisely the framework of~\cite[Section~5.2]{R}.
The recursion~(\ref{rsrec}) yields
\beq
\matrix{
R_1=1-U_1,\hfill&S_1=U_1,\hfill\cr
R_2=(1-U_1)U_2,\hfill&S_2=U_1U_2,\hfill\cr
R_3=(1-U_1)U_2U_3,\qquad\hfill&S_3=U_1U_2U_3,\hfill
}\eeq
and so on, so that
\beq
S_k=\frac{(1-R_1)R_k}{R_1}.
\eeq
The joint distribution of $R_1$ and $R_k$ is explicitly known
in the form of a double Laplace transform~\cite[Eqs.~(5.43), (5.44)]{R}.
Some algebra leads to the requested analytical expression
for the mean value $\chi=\mean{S}$:
\beq
\chi=\frac{1}{2}\int_0^\infty(1-\e^{-s})\e^{-E(s)}
\Bigl(\e^{-s}+E(s)-E(2s)\Bigr)\d s\approx0.275\,765\,063.
\label{sres}
\eeq

\begin{figure}
\begin{center}
\includegraphics[angle=-90,width=.5\linewidth]{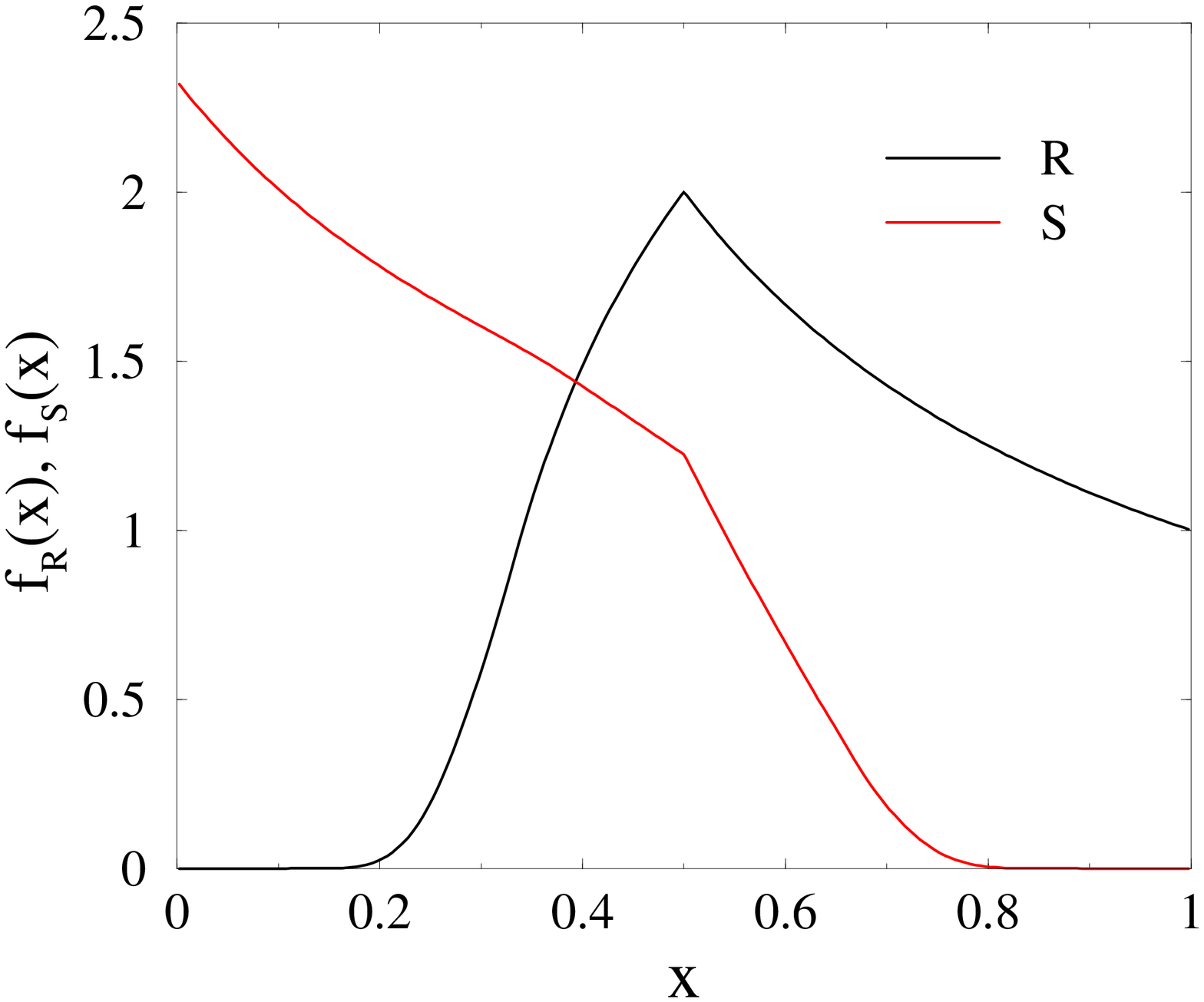}
\caption{\label{rshisto}
Densities $f_R$ and $f_S$ of the distributions
of the variables $R$ and~$S$.}
\end{center}
\end{figure}

To close up, we present in Figure~\ref{rshisto} the distributions
$f_R$ and~$f_S$, respectively giving the asymptotic distribution of the ratios
$k_\max(n)/n$ and $I(n)/n$ for the RD growth process at a generic time $n$.
These very accurate data are obtained by iterating the recursion relations~(\ref{rsrec})
$10^{10}$ times.

\subsection*{B.2.~Distribution of the ordered sequence of large degrees}

We now turn to the distribution of the whole ordered sequence of large degrees,
using the notations of Section~\ref{snapshot}.
The known relationship between the statistics of records and
that of cycles in a random permutation~\cite{golomb,goncharov,shepp,renyi,goldie}
tells us, among other things, that all the $R\lab{j}$ admit limiting distributions,
generalizing the distribution of $R\equiv R\lab{1}$ recalled above
and plotted in Figure~\ref{rshisto}.

Closed-form expressions for all the moments of all the reduced variables $R\lab{j}$
have been derived by Shepp and Lloyd~\cite{shepp}:
\beq
\mean{R\lab{j}^k}
=\int_0^\infty\frac{s^{k-1}}{k!}\,\frac{E(s)^{j-1}}{(j-1)!}\,\e^{-s-E(s)}\,\d s.
\label{rexact}
\eeq
The mean values $\mean{R\lab{j}}$
and standard deviations $\sigma(R\lab{j})=(\mean{R\lab{j}^2}-\mean{R\lab{j}}^2)^{1/2}$
of the first few variables are listed in Table~\ref{rtable}.
These numbers demonstrate the strong hierarchical ordering of the degrees
of the larger nodes.
The mean values $\mean{R\lab{j}}$ indeed fall off as~$2^{-j}$.
The number of visible large nodes therefore grows logarithmically with $n$.
We thus recover the logarithmic law of the typical number of records up to time $n$.

\begin{table}
\begin{center}
\begin{tabular}{|c|c|c|}
\hline
$j$&$\mean{R\lab{j}}$&$\sigma(R\lab{j})$\\
\hline
1&0.624\,330&0.192\,144\\
2&0.209\,581&0.112\,044\\
3&0.088\,316&0.067\,037\\
4&0.040\,342&0.039\,797\\
\hline
\end{tabular}
\end{center}
\caption{Mean value and standard deviation of the first few reduced variables $R\lab{j}$,
evaluated from the expressions~(\ref{rexact}) derived in~\cite{shepp}.}
\label{rtable}
\end{table}

The expression~(\ref{rexact}) implies that the mean value
of the moment variable $Y$ defined in~(\ref{yasy}) reads
\beq
\mean{Y}=\frac{1}{2}.
\label{yres}
\eeq
This result can be recovered alternatively as follows.
The values $Y_m$ of the moment variable at record times have been shown~\cite[Eq.~(6.6)]{R}
to obey the recursion
\beq
Y_{m+1}=U_{m+1}^2Y_m+(1-U_{m+1})^2,
\label{recy}
\eeq
with the same notations as in~(\ref{rsrec}).
Taking the mean value of both sides, we recover~(\ref{yres}).
We obtain similarly $\mean{Y^2}=7/24$, $\mean{Y^3}=139/720$, and so~on.

\end{document}